\documentclass[showpacs,twocolumn,pra,longbibliography,amsmath,amssymb,superscriptaddress,pra]{revtex4-1}

\usepackage{color}
\usepackage{graphicx}
\usepackage{amssymb}
\usepackage{amsmath}
\usepackage{mathrsfs}
\newcommand{\ud}{\mathrm{d}}

\usepackage[breaklinks=true]{hyperref}
\graphicspath{ {./figures/} {./}}

\begin{document}

\title{
Microwave generation on an optical carrier in micro-resonator chains}

\author{Andrea Armaroli}
\email{andrea.armaroli@unige.ch}
\affiliation{FOTON (CNRS-UMR 6082), Universit\'{e} de Rennes 1, ENSSAT, 6 rue de Kerampont, CS 80518, 22305 Lannion CEDEX, France}
\altaffiliation[Current address: ]{GAP-Nonlinearity and Climate, Institut of Environmental Sciences, Universit{\'e} de Gen{\`e}ve, Bd. Carl Vogt 66, 1205 Gen{\`e}ve, Switzerland}
\author{Patrice F{\'e}ron}
\affiliation{FOTON (CNRS-UMR 6082), Universit\'{e} de Rennes 1, ENSSAT, 6 rue de Kerampont, CS 80518, 22305 Lannion CEDEX, France}
\author{Yannick Dumeige}
\affiliation{FOTON (CNRS-UMR 6082), Universit\'{e} de Rennes 1, ENSSAT, 6 rue de Kerampont, CS 80518, 22305 Lannion CEDEX, France}
\date{\today}

\begin{abstract}
We consider self-pulsing regimes in chains of Kerr non-linear optical micro-resonators. By means of a super-modal diagonalization procedure of the conventional coupled-mode theory in time, we theoretically and numerically study the bifurcation diagrams of a singly-pumped 3-cavity and a doubly-pumped 4-cavity systems: the latter allows us to predict thresholdless frequency tripling of a GHz modulation. 
These self-pulsing regimes are proven robust and will find applications in generation and conversion of microwaves on an optical carrier.
\end{abstract}
\maketitle

\section{Introduction}

Optical microresonators confine light in a small volume for many optical cycles and thus provide a fundamental building block  for high speed all-optical signal processing \cite{Ilchenko2006}. Applications include frequency conversion \cite{Turner2008, Azzini2013b, Pu2015,Combrie2017}, switching \cite{Ibanescu2002,Combrie2013}, signal regeneration in communications \cite{Ghisa2007} and  optical generation of microwaves \cite{Matsko2005b,DelHaye2008,Duchesne2010}. 

One of the topics that  recently attracted the researchers' attention most is the generation of oscillations at microwave frequency by optical means. The underlying mechanism is simply the beating of optical oscillations separated by about 10 GHz to 200 GHz. This  enabling technology is necessary not only in high-speed communications (e.g.~in aerospace industry), but also in metrology, optical clocks and sensing. 
Provided that the resonator has a large enough size [so that its free spectral range (FSR) is in the target range],  a set of adjacent resonances of an optical microresonator can be used as an optical ruler \cite{Matsko2005b}. Starting from a synchronous quasi-continuous input at high enough power, the ubiquitous four-wave mixing (FWM) leads to the formation of an optical frequency comb (OFC)  \cite{Del'Haye2009,Chembo2010,Soltani2012,Cazier2013,Pasquazi2017}. This can also form train of pulses and solitary waves inside the cavity \cite{Herr2014a}. The threshold power to observe such phenomena depends on the nonlinearity of the medium and on the cavity lifetime, or equivalently the quality factor $Q$.
 This poses strong technological constraints and octave-spanning frequency combs are generally observed in  large diameter ($\approx 200 \mu \mathrm{\!m}$) glass or crystalline microresonators. Thus, due to the smallness of instantaneous optical nonlinearities, the required power levels cause thermal dissipation concerns and thus competition with thermal nonlinearities: $Q\approx10^7$ is required. Finally  in order to obtain oscillations at, e.g., 10 GHz, a microresonator of $1$ mm radius or more is needed to obtain the desired FSR. This clearly poses a serious limit to integration.
 
In order to overcome these constraints, one may rely on larger nonlinearities, such as those found in semiconductors (III-V as well as Silicon or Diamond \cite{Hausmann2014}), or on atom-like resonant effects (in bulk or low-dimensional structures, like quantum dots \cite{Grinberg2012} or nitrogen-vacancy \cite{Faraon2010a}), where, yet, the control of light-matter interactions requires cryogenic apparatuses to preserve the coherence among quantum states  \cite{Combrie2017}.

Nevertheless, an integrated nano-cavity hardly achieves $Q>10^6$, due to the unavoidable disorder and fabrication tolerances and  the FSR is in the THz range (because of its size).

Let us consider optical micro-cavities supporting a single resonant mode around the telecommunication wavelgength ($\lambda=1.55\,\mu$m). In order to trigger the system to oscillate in the  GHz range, we can couple the optical field to a microscopic  degree-of-freedom of the material, such as a time-delayed nonlinear response \cite{Malaguti2011c,Armaroli2011a,Vaerenbergh2012}. 
Beyond theoretical speculations, this proved effective for the dynamic control of photon lifetime at room temperature \cite{Huet2016}.
The coupled degrees of freedom can be also  purely optical: several cavities mutually exhanging energy. It was thought that, owing to excessive structural complexity, such a solution would be limited to two cavities \cite{Maes2009,Grigoriev2011a,Dumeige2015,Milian2018}. The disadvantage is that stable self-pulsation is limited to a period of the same order of the cavity lifetime. A short lifetime, e.g.~$0.1\,$ns, to obtain a GHz oscillation imposes an upper limit on $Q \approx 10^{4}$, thus requiring very large optical power. Moreover large injection leads to a period-doubling bifurcation cascade to chaos.
In \cite{Armaroli2015cav}, we showed that  a system of three evanescently-coupled optical micro-race-track resonators with instantaneous Kerr response can be tuned to oscillate at a frequency that depends only on the coupling between the cavities (supposed large). This solutions is compatible with a $Q\approx10^5$, so it cuts down the injection energy requirements;
our approach is more robust and flexible.

A simple qualitative description of the mechanism goes as follows: thanks to large coupling, the three resonant frequencies (that are assumed coincident for isolated resonators) split far apart, much more than the detuning between the central uncoupled resonance  and the laser injection. The resulting  {collective modes of the whole structure, denoted hereafter as \emph{super-modes}} are coupled by means of  degenerate FWM. The energy is transfered from the (externally excited) central mode to lateral ones, similarly to frequency combs or a multi-modal nano-cavity \cite{Belotti2010a,Combrie2017}, and may result in stable oscillations.

Inspired by similar studies in the theory of frequency combs \cite{Hansson2013}, in the present work we develop a more detailed analysis of the three-cavity oscillator and present its bifurcation diagram. After having assessed the validity of this method, we extend it to a doubly pumped four-cavity chain, analogous to bichromatically pumped frequency combs \cite{Strekalov2009,Hansson2014d}, which are based on non-degenerate FWM. This is a non-autonomous set of ordinary differential equations (ODEs), which is greatly simplified by our  method.

We prove that in the large coupling coefficient regime, a microwave frequency $\Omega$ can be converted with high efficiency to $3\Omega$---up to $50\%$ of the total energy inside stored in the cavity system, i.e.~half of the energy coupled from an external waveguide is used to excite super-modes at $\pm\Omega$ and half at $\pm 3\Omega$. The bifurcation diagram is similar to that of the three-cavity system, but, owing to the threshold-less nature of non-degenerate FWM, no minimum power is required. 

Contrary to bichromatically pumped combs, where FWM is further cascaded over a broad bandwidth and a truncated dynamics behaves appropriately only for quite a specific range of parameters, our treatment is consistent and robust: it can be extended to chains (or molecules) of an arbitrary number of coupled resonators. 

In section \ref{sec:CSMT} we introduce the  diagonalization procedure of the nonlinear time-dependent coupled-mode equations (CMT)  \cite{HausBook} to coupled super-mode equations (CSMT). In \ref{sec:3cav} we apply this to revisit and improve our understanding of the results of \cite{Armaroli2015cav}. Remarkably, we present a new phase-space representation and a more detailed bifurcation diagram. In section \ref{sec:4cav} we study the four cavity case and contrast its results to that of the previous section. In section \ref{sec:implementation}, we discuss the accessibility and technological feasibility of the proposed solutions, by means of a Monte-Carlo approach to explore the parameter space. Finally, in section \ref{sec:conclusion}, we conclude.

\section{Coupled-modes and coupled-supermodes}
\label{sec:CSMT}

Let us consider a system composed by $M$ evanescently coupled optical microcavities (single mode or with a large FSR), the time evolution of which reads, in dimensional units, as \cite{Fan2003,Grigoriev2011a,Abdollahi2014,Dumeige2015}
\begin{equation}
\begin{split}
\label{eq:dim}
\frac{\ud{A}_j}{\ud T}=\left[i(\tilde\delta_j+\tilde\chi_j\left|A_j\right|^2)-\frac{1}{\tilde \tau_j}\right]A_j \\+i \sum_{k\neq j}{\tilde\gamma_{jk} A_k}+ \sqrt{\frac{2}{\tilde\tau_\mathrm{wg}}} s_\mathrm{in}(T)\delta^K_{j1}
\end{split}
\end{equation}
where $j,k=1,\ldots,M$, $\tilde\delta_j=\tilde\omega_j-\tilde\omega_L$ is the detuning of the laser excitation from the $j$-th cavity resonance frequency, $\tilde \tau_j$ is the cavity lifetime, $\tilde\tau_\mathrm{wg}$ quantifies the coupling from the input waveguide to the first cavity. $\delta^K_{mn}$ is the Kronecker delta. We further assume, for the sake of simplicity, that the decay into the waveguide is negligible with respect to the intrinsic cavity contribution, i.e.~$\tilde \tau_1\ll\tilde\tau_\mathrm{wg}$  (undercoupling), as opposed to the  critically coupled (the escape and decay rates are equal) or overcoupled (the escape in the waveguide is dominant) case \cite{Alphonse2014}. 

$A_j$ are normalized such that  $|A_j|^2$ is the energy stored in the cavity (in J), $|s_{in}|^2$ is the power in W in the external waveguide coupled into the first cavity, $\tilde\gamma_{jk}=\tilde\gamma_{kj}$ are the coupling rate of cavity $j$ and $k$ and are assumed to be real. The independent variable $T$ represents the time expressed in seconds.
We let the effective nonlinear coefficient $\tilde\chi_j=\frac{\tilde\omega_j c n_2}{n_\mathrm{eff}^2\mathcal{V}}$, where $n_2$ is the Kerr coefficient, $n_\mathrm{eff}$ is the modal effective index and $\mathcal{V}$ is the modal effective volume. 

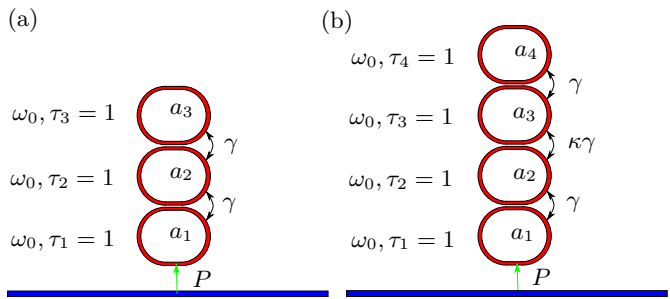
\begin{figure}[ht]
	\centering
					\def\svgwidth{.49\textwidth}
\begingroup%
  \makeatletter%
  \providecommand\color[2][]{%
    \errmessage{(Inkscape) Color is used for the text in Inkscape, but the package 'color.sty' is not loaded}%
    \renewcommand\color[2][]{}%
  }%
  \providecommand\transparent[1]{%
    \errmessage{(Inkscape) Transparency is used (non-zero) for the text in Inkscape, but the package 'transparent.sty' is not loaded}%
    \renewcommand\transparent[1]{}%
  }%
  \providecommand\rotatebox[2]{#2}%
  \ifx\svgwidth\undefined%
    \setlength{\unitlength}{1013.17059855bp}%
    \ifx\svgscale\undefined%
      \relax%
    \else%
      \setlength{\unitlength}{\unitlength * \real{\svgscale}}%
    \fi%
  \else%
    \setlength{\unitlength}{\svgwidth}%
  \fi%
  \global\let\svgwidth\undefined%
  \global\let\svgscale\undefined%
  \makeatother%
  \begin{picture}(1,0.43249811)%
    \put(0,0){\includegraphics[width=\unitlength]{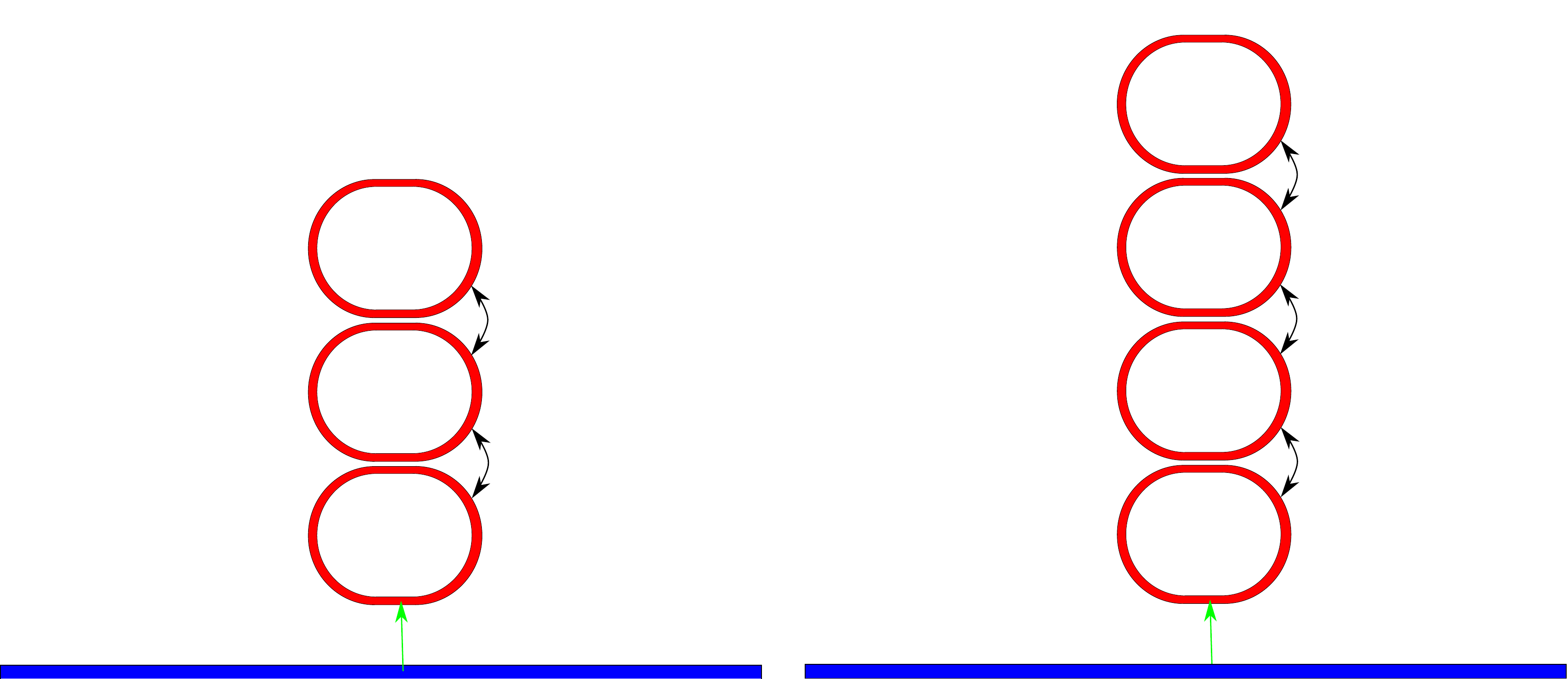}}%
    \put(0.21018262,0.50112973){\color[rgb]{0,0,0}\makebox(0,0)[lt]{\begin{minipage}{0.04890872\unitlength}\raggedright \end{minipage}}}%
    \put(0.11955765,0.45977309){\color[rgb]{0,0,0}\makebox(0,0)[lt]{\begin{minipage}{0.0406374\unitlength}\raggedright \end{minipage}}}%
    \put(0.27301178,0.12531996){\color[rgb]{0,0,0}\makebox(0,0)[lt]{\begin{minipage}{0.10377607\unitlength}\raggedright \end{minipage}}}%
    \put(0.3615502,0.58096255){\color[rgb]{0,0,0}\makebox(0,0)[lt]{\begin{minipage}{0.10049972\unitlength}\raggedright \end{minipage}}}%
    \put(0.29410059,0.6463168){\color[rgb]{0,0,0}\makebox(0,0)[lt]{\begin{minipage}{0.04890872\unitlength}\raggedright \end{minipage}}}%
    \put(0.20347561,0.60496016){\color[rgb]{0,0,0}\makebox(0,0)[lt]{\begin{minipage}{0.04063739\unitlength}\raggedright \end{minipage}}}%
    \put(0.35692975,0.27050703){\color[rgb]{0,0,0}\makebox(0,0)[lt]{\begin{minipage}{0.10377607\unitlength}\raggedright \end{minipage}}}%
    \put(0.3615502,0.27308244){\color[rgb]{0,0,0}\makebox(0,0)[lb]{\smash{}}}%
    \put(0.3144908,0.60568869){\color[rgb]{0,0,0}\makebox(0,0)[lb]{\smash{}}}%
    \put(0.09298289,0.39745724){\color[rgb]{0,0,0}\makebox(0,0)[lt]{\begin{minipage}{0.06655204\unitlength}\raggedright \end{minipage}}}%
    \put(0.12795091,0.43411725){\color[rgb]{0,0,0}\makebox(0,0)[lt]{\begin{minipage}{0.000564\unitlength}\raggedright \end{minipage}}}%
    \put(0.84838689,0.24031932){\color[rgb]{0,0,0}\makebox(0,0)[lt]{\begin{minipage}{0.26161836\unitlength}\raggedright $\kappa \gamma$\end{minipage}}}%
    \put(0.84653573,0.15175196){\color[rgb]{0,0,0}\makebox(0,0)[lt]{\begin{minipage}{0.26161836\unitlength}\raggedright $\gamma$\end{minipage}}}%
    \put(0.79418149,0.0423366){\color[rgb]{0,0,0}\makebox(0,0)[lt]{\begin{minipage}{0.03622446\unitlength}\raggedright $P$\end{minipage}}}%
    \put(0.51807453,0.28453489){\color[rgb]{0,0,0}\makebox(0,0)[lt]{\begin{minipage}{0.15843696\unitlength}\raggedright $\omega_0,\tau_3=1$\end{minipage}}}%
    \put(0.51761902,0.190183){\color[rgb]{0,0,0}\makebox(0,0)[lt]{\begin{minipage}{0.17325376\unitlength}\raggedright $\omega_0,\tau_2=1$\end{minipage}}}%
    \put(0.5209388,0.09878659){\color[rgb]{0,0,0}\makebox(0,0)[lt]{\begin{minipage}{0.17023729\unitlength}\raggedright $\omega_0,\tau_1=1$\end{minipage}}}%
    \put(0.76218316,0.0875406){\color[rgb]{0,0,0}\makebox(0,0)[lb]{\smash{$a_1$}}}%
    \put(0.90202287,0.37427381){\color[rgb]{0,0,0}\makebox(0,0)[lb]{\smash{ }}}%
    \put(0.76518671,0.27036809){\color[rgb]{0,0,0}\makebox(0,0)[lb]{\smash{$a_3$}}}%
    \put(0.76557776,0.17880663){\color[rgb]{0,0,0}\makebox(0,0)[lb]{\smash{$a_2$}}}%
    \put(0.84998944,0.32982445){\color[rgb]{0,0,0}\makebox(0,0)[lt]{\begin{minipage}{0.26161836\unitlength}\raggedright $\gamma$\end{minipage}}}%
    \put(0.52120834,0.37527713){\color[rgb]{0,0,0}\makebox(0,0)[lt]{\begin{minipage}{0.20290518\unitlength}\raggedright $\omega_0,\tau_4=1$\end{minipage}}}%
    \put(0.76780281,0.36417841){\color[rgb]{0,0,0}\makebox(0,0)[lb]{\smash{$a_4$}}}%
    \put(0.47448041,0.43441267){\color[rgb]{0,0,0}\makebox(0,0)[lt]{\begin{minipage}{0.05583318\unitlength}\raggedright (b)\end{minipage}}}%
    \put(0.00022749,0.43731721){\color[rgb]{0,0,0}\makebox(0,0)[lt]{\begin{minipage}{0.05583318\unitlength}\raggedright (a)\end{minipage}}}%
    \put(0.32807905,0.23880507){\color[rgb]{0,0,0}\makebox(0,0)[lt]{\begin{minipage}{0.26161836\unitlength}\raggedright $\gamma$\end{minipage}}}%
    \put(0.32555089,0.15003685){\color[rgb]{0,0,0}\makebox(0,0)[lt]{\begin{minipage}{0.26161836\unitlength}\raggedright $\gamma$\end{minipage}}}%
    \put(0.28052874,0.03872146){\color[rgb]{0,0,0}\makebox(0,0)[lt]{\begin{minipage}{0.03592511\unitlength}\raggedright $P$\end{minipage}}}%
    \put(0.24540517,0.08841984){\color[rgb]{0,0,0}\makebox(0,0)[lb]{\smash{$a_1$}}}%
    \put(0.24598135,0.2765068){\color[rgb]{0,0,0}\makebox(0,0)[lb]{\smash{$a_3$}}}%
    \put(0.24579378,0.18031836){\color[rgb]{0,0,0}\makebox(0,0)[lb]{\smash{$a_2$}}}%
    \put(0.00782619,0.29094984){\color[rgb]{0,0,0}\makebox(0,0)[lt]{\begin{minipage}{0.19777723\unitlength}\raggedright $\omega_0,\tau_3=1$\end{minipage}}}%
    \put(0.00503216,0.19195717){\color[rgb]{0,0,0}\makebox(0,0)[lt]{\begin{minipage}{0.25101865\unitlength}\raggedright $\omega_0,\tau_2=1$\end{minipage}}}%
    \put(0.00607703,0.10025007){\color[rgb]{0,0,0}\makebox(0,0)[lt]{\begin{minipage}{0.23559673\unitlength}\raggedright $\omega_0,\tau_1=1$\end{minipage}}}%
  \end{picture}%
\endgroup%
					\caption{(Color Online) The two different chains of nonlinear resonators under study. (a) Three-cavity configuration; (b) four-cavity configuration, notice that the central coupling coefficient allows us to tune the super-mode frequency spacing.}
	\label{fig:scheme}
\end{figure}

As in \cite{Armaroli2015cav}, we assume that the lifetimes and modal properties are the same for each cavity $\tilde\tau_j=\tilde\tau$, $\tilde\omega_j=\tilde\omega$, $\tilde\delta_j=\tilde\delta$, $\tilde\chi_j=\tilde\chi>0$ for $j=1, \dots, M$. As for the coupling coefficients, we consider a linear chain of resonators (as in Fig.~\ref{fig:scheme}). We thus limit ourselves to $\tilde\gamma_{jk}\neq 0$ for $|j-k|=1$, $j,k>0$, and $\tilde\gamma_{jk}=0$ otherwise. We will define below what is the most symmetric choice for the different situations.

By introducing the normalization $a=A/\sqrt{I_0}$, $t=T/\tilde\tau$, with $I_0=(\tilde\tau \tilde\chi)^{-1}$, we derive from \eqref{eq:dim} the following adimensional model
\begin{equation}
\begin{aligned}
\dot{a}_j&=\left[i(\delta+\chi\left|a_j\right|^2)-1\right]a_j +i\left( \gamma_{j,j-1} a_{j-1} + \gamma_{j,j+1} a_{j+1}\right) \\&+ \sqrt{P}f(t)\delta^K_{j1}
\end{aligned}
\label{eq:adim}
\end{equation} 
where the dot denotes the derivative in $t$; 
moreover $\gamma_{jk}=\tilde\gamma_{jk}\tilde\tau$, $\delta=\tilde\delta\tilde\tau$ and $P=2\tilde\tau^2 |s_{in}|^2/(I_0\tilde\tau_\mathrm{wg})=2\tilde\tau^3 |\tilde\chi||s_{in}|^2/\tilde\tau_\mathrm{wg}$ is the actual power coupled in the first cavity.   $f(t)$ is a normalized function which denotes the envelope of the input signal coupled from the waveguide. 

If $f(t)=1$, the conventional approach consists in finding the equilibria (fixed points) of \eqref{eq:adim}  by imposing $\ud a_j/\ud t=0$, for $j=1,\dots,M$, and characterize their properties. 
We proved in \cite{Armaroli2015cav} that, if the coupling coefficients are large compared to the cavity lifetime, $\gamma_{jk}\gg 1$, the dynamics greatly simplifies and regular self-pulsing regimes are found. They can be thought as beating between super-modes. 

However this is not possible if the system is not autonomous, i.e.~if $f(t)$ has an explicit time-dependence. This is the case, for example, of an harmonically forced system, see below. We employ an harmonic balance technique; alternative approaches for computing bifurcation curves  are possible \cite{Sracic2011}, the most common is to include several coupled oscillating degrees of freedom to transform it to  an autonomous form.

In details, let $\mathbf{a} \equiv [a_1,\dots,a_M]^T$ the vector of complex modal amplitudes, we write Eq.~\eqref{eq:adim} in compact form
\begin{equation}
\dot{\mathbf{a}} = (i\delta-1) \mathbf{a}  + i\mathbf{\Gamma} \mathbf{a} + \chi \mathbf{N}(\mathbf{a}) + \sqrt{P} f(t) \mathbf{\delta_1^K}
\label{eq:VCMT} 
\end{equation}
where the definitions of $\mathbf{\Gamma}$, $\mathbf{N}(\mathbf{a})$, and $\mathbf{\delta_1^K}$ are obvious,  by comparing Eq.~\eqref{eq:adim} to Eq.~\eqref{eq:VCMT}.

We suppose that the non-linear part is small and the solution is a perturbation of the linear solution.

We start from the homogeneous linear system obtained  for $\chi=0$, $P=0$, and $\dot{\mathbf{a}}=0$, which provides the collective modes (super-modes) of the system. The  complex resonant frequencies (eigenvalues) $\Omega_j^C \equiv \Omega_j+i$, with $j=1,\ldots,M$ (i.e.~the lifetime is the same as for isolated cavities, provided that every cavity has the same resonant frequency).
Notice that we diagonalize only the coupling part, i.e.~$\mathbf{\Gamma = V \Delta V^{-1}}$, with $\mathbf{\Delta}=\mathrm{diag}\,[\Omega_1,\Omega_2,\ldots,\Omega_M]$ the diagonal matrix of eigenvalues and $\mathbf{V}$ the matrix of eigenvectors. 

By defining the super-mode complex amplitude as 
$\mathbf{a} = \mathbf{V}\exp(i\mathbf{\Delta} t) \mathbf{u}$, we obtain a system of coupled non-linear equations in the form
\begin{equation}
\begin{aligned}
\dot{\mathbf{u}}&=(i\delta-1)\mathbf{u}+ \exp(-i\Delta t)\mathbf{V}^{-1}\mathbf{N}(\mathbf{V}\exp(i\Delta t)\mathbf{u})\\
&+\exp(-i\Delta t)\mathbf{V}^{-1}  \sqrt{P} f(t) \mathbf{\delta_1^K},
\end{aligned}
\label{eq:CSMT}
\end{equation}
with no approximations with respect to 
Eq.~\eqref{eq:VCMT}. 
 {Further simplifications can be made by neglecting all oscillating terms in Eq.~\eqref{eq:CSMT}: thus only phase-matched non-linear terms and constant forcing are retained. This is possible if $f(t)$ is a sum of sinusoids with frequencies equal to linear combinations of $\Omega_j$ with integer coefficients and allows to transform a non-autonomous system into an autonomous one.}

In this work we consider a three-cavity and a  four-cavity system, respectively in Secs.~\ref{sec:3cav} and \ref{sec:4cav}, as prototypes of degenerate- and non-degenerate four-wave mixing (FWM) among super-modes. 
Our approach allows us to derive an autonomous system, the fixed points of which are easily characterized.

We will highlight their analogies and differences by analyzing their full nonlinear evolution.

\section{Three-cavities revisited}
\label{sec:3cav}
\subsection{Derivation of coupled-super-modes theory}
We start by revisiting the results of Ref.~\citep{Armaroli2015cav}.
Consider the $M=3$ system depicted in Fig.~\ref{fig:scheme}(a). The coupling coefficients are supposed identical, $\gamma_{12}=\gamma_{23}=\gamma$, and $f(t)=1$. This highly symmetric configuration supports three super-modes at $\Omega_0=0$ and $\Omega_{\mp 1}=\mp \sqrt{2}\gamma$, with unit normalized lifetime. The eigenvector matrix reads as
\[
	\mathbf{V}=
	\begin{bmatrix}
		\frac{\sqrt{2}}{2} & \frac{1}{2} & \frac{1}{2}\\
		0 & -\frac{\sqrt{2}}{2} & \frac{\sqrt{2}}{2}\\
		-\frac{\sqrt{2}}{2} & \frac{1}{2} & \frac{1}{2}\\
	\end{bmatrix}.
\]
 {The amplitudes $\mathbf{u}=[u_0,u_{-1},u_{1}]^T$ of the super-modes evolve according to Eq.~\eqref{eq:CSMT}. If we  neglect all oscillating terms, we write}
\begin{equation}
		\begin{aligned}
		\dot{u}_0 =& (i\delta-1)u_0  + \sqrt{\frac{P}{2}}
			\\&+  i\frac{\chi}{2}\left[\left(\underbrace{|u_0|^2}_\mathrm{SPM}+\underbrace{|u_{-1}|^2 + |u_{1}|^2}_\mathrm{XPM}\right){u_0}+\underbrace{u_0^* u_{-1} u_{1}}_\mathrm{Coher.}\right]			
   \\   
   			\dot{u}_{-1} =& (i\delta-1)u_{-1}  
   			\\&+i\frac{\chi}{8}\left[\left(\underbrace{3|u_{-1}|^2}_\mathrm{SPM} + \underbrace{4|u_0|^2 + 6|u_{1}|^2}_\mathrm{XPM}\right)u_{-1} + \underbrace{2u_{1}^* u_0^2}_\mathrm{Coher.}
   			\right]
   			\\
   			\dot{u}_1 =& (i\delta-1)u_{1} 
   			\\&+ i \frac{\chi}{8}\left[\left(\underbrace{3|u_{1}|^2}_\mathrm{SPM} + \underbrace{4|u_0|^2+6|u_{-1}|^2}_\mathrm{XPM}\right)u_{1}+
   				\underbrace{2{u_{-1}^* u_0^2}}_\mathrm{Coher.}\right]
		\end{aligned}
		\label{eq:CSMT3}
\end{equation}	
Each super-mode undergoes equal dephasing and loss, owing to the equality of lifetimes and resonant frequencies; $u_0$ is coherently pumped by the external waveguide. The nonlinear response is composed of a self-phase modulation (SPM), a cross-phase modulation (XPM), and a coherent transfer of energy from one mode to the others. While this form  is easily predictable by virtue of the cubic nonlinearity, the coefficients differ from similar systems \cite{Chembo2010}. Notice that $\gamma$ does not appear anymore.  {In Eq.~\eqref{eq:CSMT3} we neglect: (i) phase-mismatched non-linear terms oscillating at linear combination of the super-mode frequencies different from $0$  (ii) Forcing terms oscillating at $\Omega_{\mp1}$. The full forcing vector reads as $[\sqrt{\frac{P}{2}},\frac{\sqrt{P}}{2}\exp{(i\Omega_{1} t)},\frac{\sqrt{P}}{2}\exp{(i\Omega_{-1} t)}]^\mathrm{T}$ and is the main responsible to the small oscillations we observe in Figs.~\ref{fig:three_evol1} and \ref{fig:three_evol2} below.}

The stationary response of the system is calculated by letting $u_{\pm 1}=0$. We can write
$P=2\eta_0\left[1+\left(\delta+\frac{\chi}{2}\eta_0\right)^2\right]$, where $\eta_0 \equiv |u_0|^2$. 
The system is thus bistable if $\delta<-\sqrt{3}$, and the two saddle-node bifurcations occur for $\frac{\chi \eta_0^\pm}{2} = -\frac{2}{3}\delta \pm \frac{\sqrt{\delta^2-3}}{3}$.

The self-pulsing threshold is derived by assuming $|u_{\pm 1}|\ll |u_0|$ so that no appreciable reverse conversion of energy from sidebands to carrier is possible. The linearized system reads as
\begin{align*}
	\dot{u}_{-1} =& (i\delta-1)u_{-1}  
   			+i\frac{\chi}{4}\left[2|u_0|^2u_{-1}+u_1^* u_0^2
   			\right]
   			\\
   			\dot{u}_1 =& (i\delta-1)u_{1} 
   			+ i \frac{\chi}{4}\left[2|u_0|^2u_{1}+
   				{u_{-1}^* u_0^2}\right]
\end{align*}
and predicts growing sidebands (thus the initiation of self-pulsing) if the gain $G\equiv   \sqrt{-\delta^2 - \chi\delta \eta_0 - \frac{3\chi^2 \eta_0^2}{16}}$ overcomes losses ($G>1$), i.e.~if $\eta_0^{H-}<\eta_0<\eta_0^{H+}$, with $\frac{\chi \eta_0^{H\pm}}{4} = -\frac{2}{3}\delta \pm \frac{\sqrt{\delta^2-3}}{3}$. These results correspond to Eqs.~(4-5) of  \cite{Armaroli2015cav}, by observing that $I_3=I_1=\eta_0/2$, letting $\gamma\to \infty$.

A more detailed analysis of the limit cycles is also possible: Eq.~\eqref{eq:CSMT3} has the same form of the coupled-mode theory derived in \cite{Hansson2013}, apart from the different weights of SPM and XPM.
On the lines of that work,
in order to precisely describe the equilibria of Eq.~\eqref{eq:CSMT3}, we transform it to real variables. We define $u_k = \sqrt{\eta_k}\exp (i \phi_k)$, for $k=-1,0,1$, and notice from  Eq.~\eqref{eq:CSMT3} (and confirm numerically below, in Figs.~\ref{fig:three_evol1} and \ref{fig:three_evol2}) that the steady-state sideband imbalance $\alpha\equiv\eta_{1}-\eta_{-1}$ is a constant of motion. It is verified that $\alpha=0$. Thus, only four real variables are required, i.e.~the relative phase $\psi \equiv 2 (\phi_1 -\phi_0)$, the relative pump-central-mode phase, $\phi_0$ (Eq.~\eqref{eq:CSMT3} is not phase invariant), the total intensity in the modes, $E \equiv \eta_0 + 2\eta_{1}$, and the fraction of the intensity in the lateral modes, $\eta\equiv (1-\eta_0)/E$. We derive, after some simple algebra, 
\begin{equation}
\begin{aligned}	
	\dot\eta &= \eta(1-\eta)\left(\frac{\chi E}{2}\sin \psi - \sqrt{\frac{2P}{(1-\eta) E}}\cos\phi_1\right)\\
	\dot E &= -2E +\sqrt{2(1-\eta) E P }\cos \phi_1\\
	\dot \psi &= \sqrt{\frac{2P}{(1-\eta) E} }\sin \phi_1 
	+\frac{\chi E}{2}(1-2\eta) \cos \psi + \frac{\chi E}{8}\eta\\
	\dot \phi_1 & =\left(\delta + \frac{\chi E }{2}\right)
	-\sqrt{\frac{P}{2(1-\eta) E}}\sin\phi_1 + 
	\frac{\chi E}{4}\eta \cos \psi\\	
\end{aligned}
\label{eq:realCSMT3}
\end{equation}
The bistable response mentioned above corresponds to $\eta=0$; no fixed point is consistent with $\eta=1$. The non-trivial fixed points ($\eta\neq 0$)
can be expressed in implicit form, by means of some cumbersome algebra. $\eta E$ represents directly the energy in the sidebands, i.e.~the generated microwave signal.  

\subsection{Bifurcation diagram and numerical results}
\begin{figure}[hbtp]
\centering
\includegraphics[width=.45\textwidth]{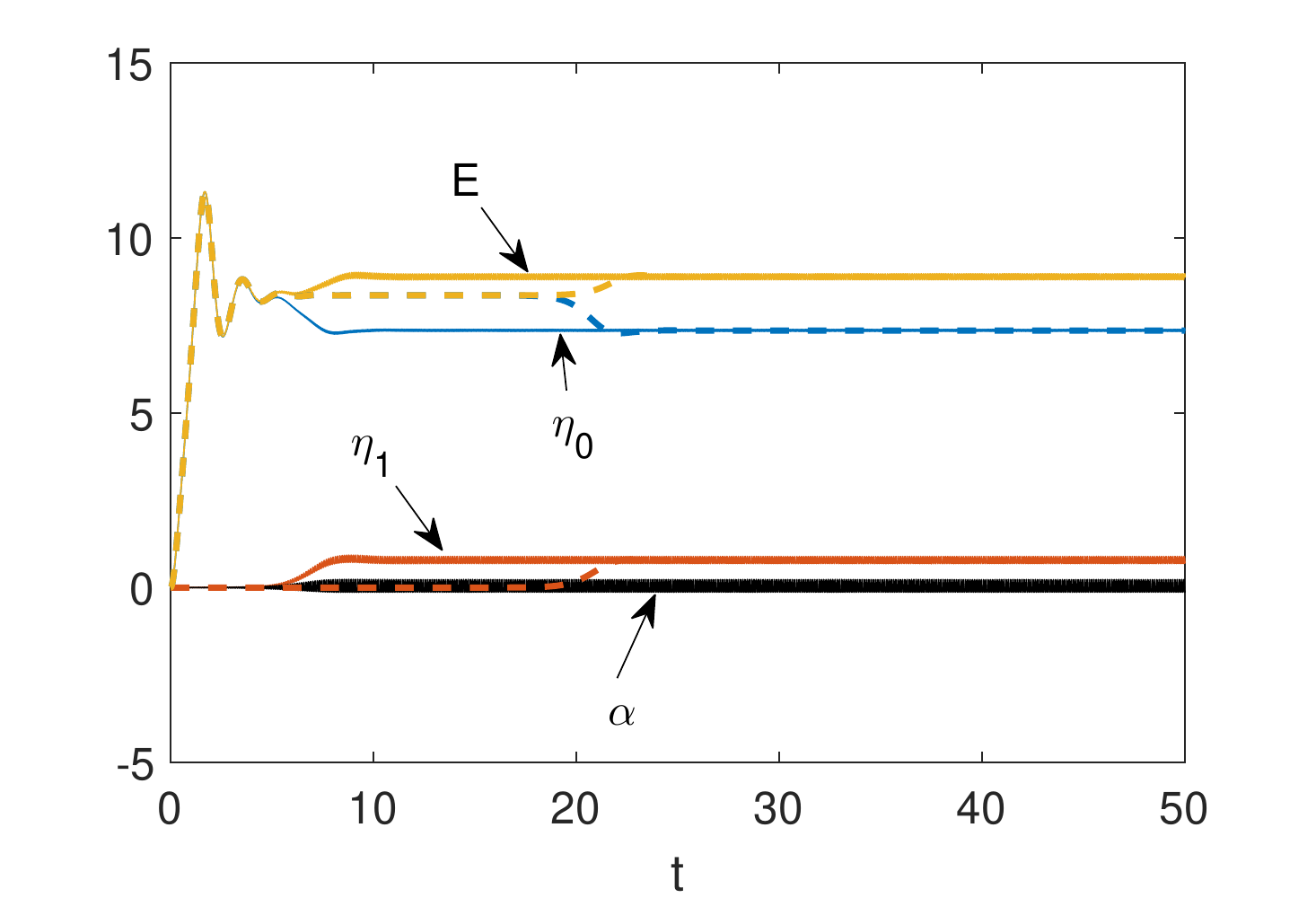}
\caption{(Color Online) Comparison of the evolution toward steady-state of the three cavity system and its three mode truncation. $P=40$, $\delta=-3$, $\gamma=40$. Solid lines represent the solution of Eq.~\eqref{eq:adim}, while dashed lines the solution of Eq.~\eqref{eq:CSMT3}: $\eta_0$ is in blue (dark grey upper line), $\eta_1$ in red (grey at the bottom), $E$ in yellow (light gray at the top), the imbalance $\alpha$ is finally in black (at the bottom) [only for Eq.~\eqref{eq:adim}]. Notice that the last one exhibits small oscillations around zero. The other quantities reach steady-state more slowly in the Eq.~\eqref{eq:CSMT3} than in the original model, but represent quite an accurate approximation of the system behavior.}
\label{fig:three_evol1}
\end{figure}
In Fig.~\ref{fig:three_evol1}, we compare the results of the present approach a case qualitatively similar to what we presented in \cite{Armaroli2015cav} [Fig. 5(b)]. We let $\gamma=40$, $P=40$, $\delta=-3$ and study the evolution in time from noisy initial conditions of the cavity towards its self-pulsing state [the slight increase in $P$ (from 30 in our previous study to 40) allows us to show more clearly the different bifurcations]. We plot the intensities of the super-modes and  observe that the time  at which the steady-state is achieved is different from Eq.~\eqref{eq:adim} to \eqref{eq:CSMT3}, but the  values at which $\eta_0$ and $\eta_{\pm 1}$ stabilize are close and the behavior is qualitatively similar: importantly the energy transferred to sidebands, i.e.~the microwave output, is particularly well-predicted. The imbalance $\alpha$ is oscillating about zero, thus reassuring us on the soundness of the present approximation. 

\begin{figure}[hbtp]
\centering
\includegraphics[width=.45\textwidth]{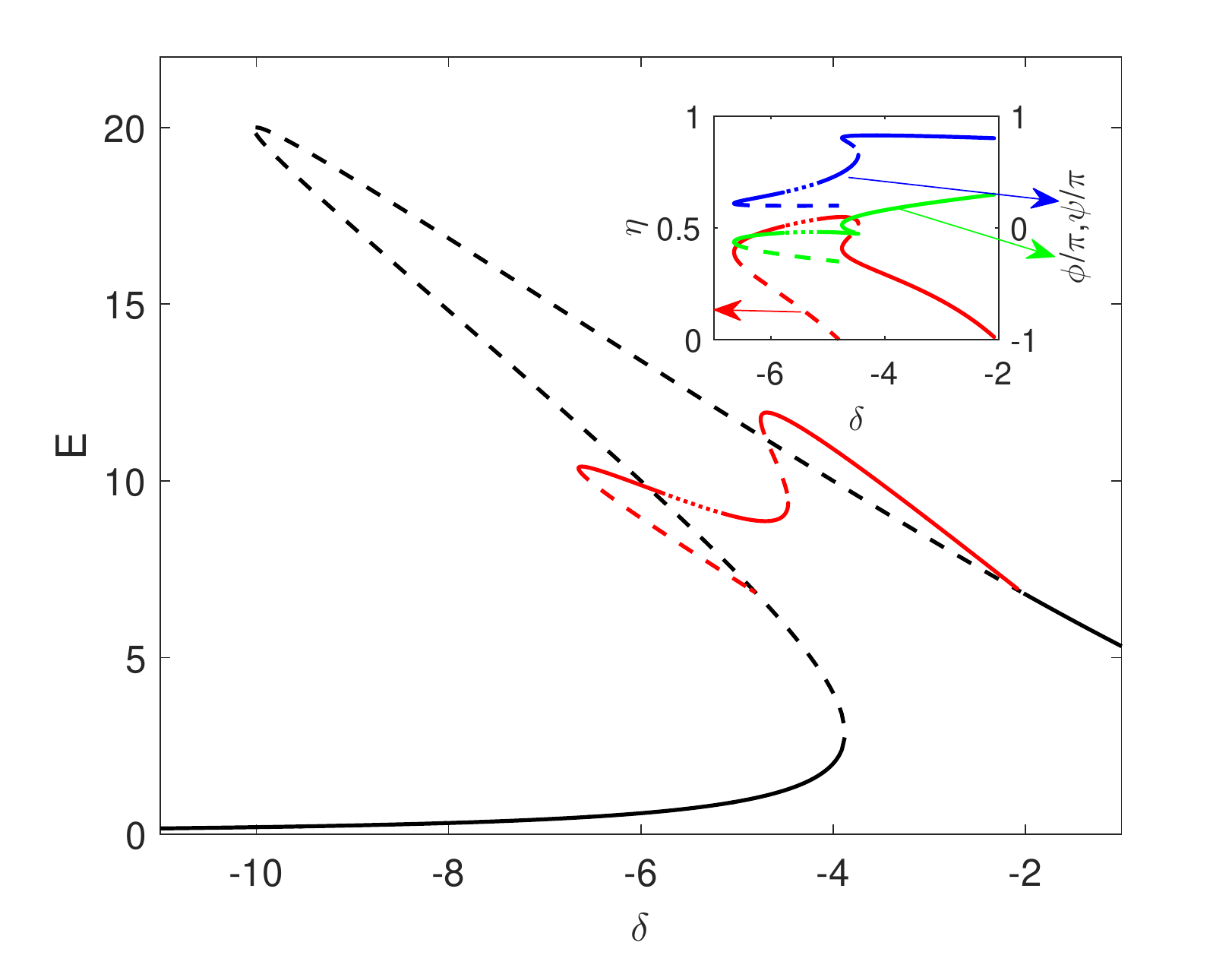}\caption{(Color Online) Bifurcation diagram with $P=40$. The main panel shows the bistable curve of $E$ as a function of $\delta$ (bifurcation parameter) of fixed points of Eq.~\eqref{eq:CSMT3}, with $\eta=1$, in black, and limit cycles, $\eta\neq 0,1$, in red (gray). Solid  lines represent stable  limit cycles, dashed lines unstable (saddle points) and dotted lines are delimited by Neimark-Sacker  bifurcations. The inset shows the bifurcation of the other three variables, $\eta$ [red (gray), left axis], $\phi_1$ [green (light gray), right axis], and $\psi$ [blue (dark gray), right axis].  }
\label{fig:bifurcation3}
\end{figure}
Fig.~\ref{fig:bifurcation3} shows the bifurcation diagram  for the trivial (bistable curve of fixed points, in black) and non-trivial solutions  [limit cycles, in red (gray)] of Eq.~\eqref{eq:realCSMT3} obtained by means of a standard numerical continuation package \cite{Dhooge2003}. This is equivalent to Fig.~4 of our previous work, we simply change the observables that we consider: $E$ instead of the energy stored into the third cavity. The inset shows the fraction $\eta$ in the lateral modes and the relative phases $\psi$ and $\phi_1$, which were not straightforward to obtain from our previous calculations. The results of the main panel are  indistinguishable from the results of the  bifurcation analysis presented in \cite{Armaroli2015cav}, once adapting them to the new variable: in the main panel the comparison is irrelevant. 
As we know well two types of instability coexist: the absolute instability which leads to the conventional bistable response and the Andronov-Hopf bifurcation which destabilizes the upper stable state and corresponds to the initiation of self pulsing. The bifurcation diagram of limit cycles [in red (gray)] exhibits in turn a saddle-node bifurcation. 
While the branch of limit cycles for $-4.75<\delta<-2.08$ corresponds to a larger enervgy stored in the cavity than in the unstable equilibria, for lower $\delta$, $E$ decreases and  $\eta$ stabilizes after a sudden surge (red line in the inset). For $\delta<-3.88$ the system is attracted to the lower fixed point. Another branch of stable limit cycles [separated by a short branch of unstable solutions, dashed lines in the inset and red (gray) dashed line in the main panel] exists for $-6.65<\delta<-4.47$, but it can be reached only from a hot cavity state, i.e.~non-zero initial mode amplitudes. limit cycles in this branch undergo themselves a bifurcation, namely a Neimark-Sacker (NS, i.e.~secondary Andronov-Hopf) bifurcation, from a cycle to a torus, as discussed below. Notice that the conversion efficiency $\eta>0.5$ for $-5.12<\delta<-4.47$, the central mode contains less energy than the sidebands.

Finally, the  limit cycles reconnect to the family of fixed points via a branch of unstable solutions. No other states separated from the curve of fixed points where found, in contrast to \cite{Hansson2013}. If the interval between the two NS points is small enough the most-detuned segment of the second branch can be reached adiabatically from the first one. We can thus state that our system is a soft-excitable self-pulsating system. 

\begin{figure}[hbtp]
\centering
\includegraphics[width=.49\textwidth]{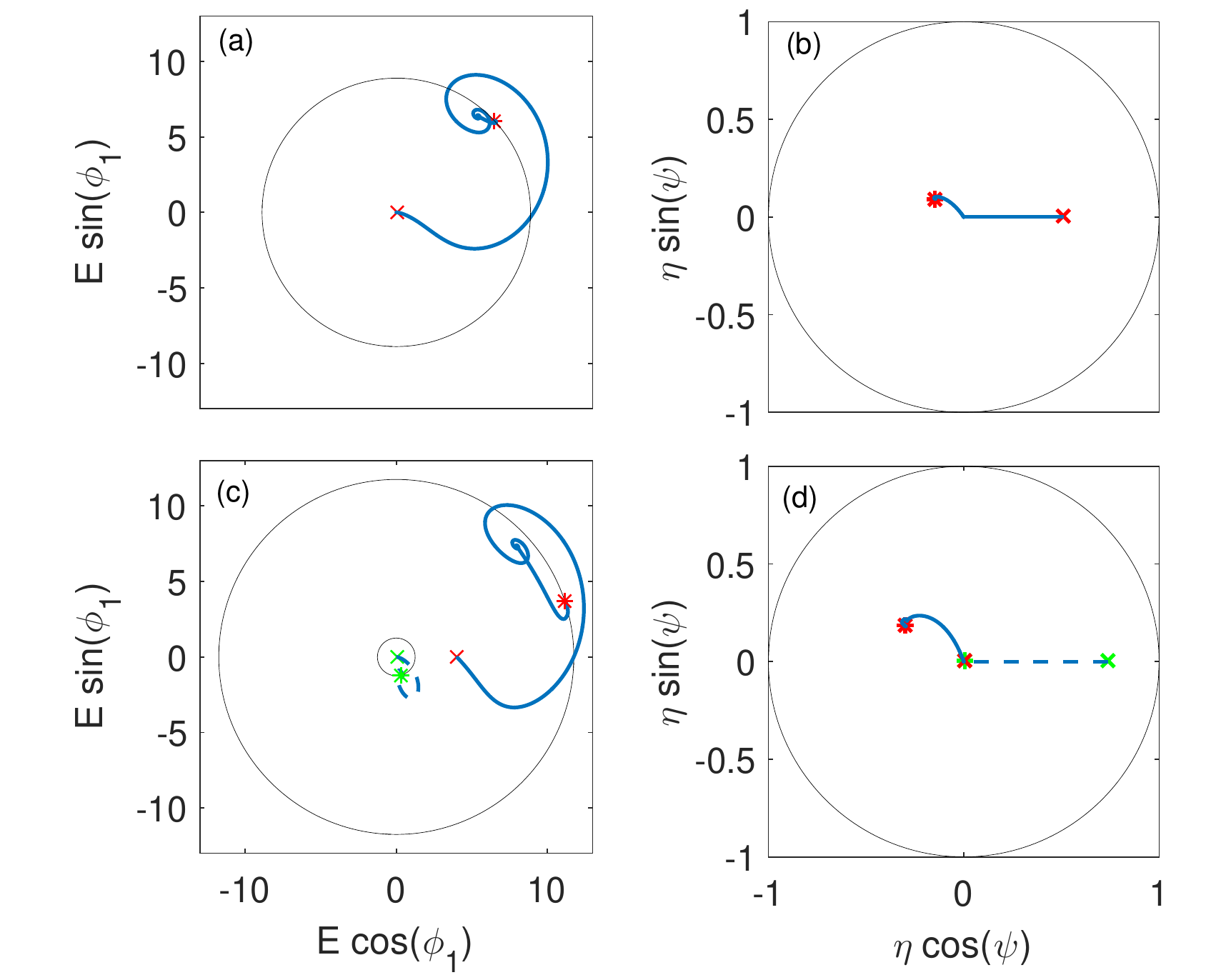}
\caption{(Color Online) Phase space representations of the time evolution of Eq.~\eqref{eq:CSMT3}. (a) and (c) show the plane $(E\cos\phi_1, E\sin\phi_1)$, the circles denoting steady $E$s, (b) and (d) the plane $(\eta\cos\psi, \eta\sin\psi)$, the unit circles included for reference.  (a-b) Correspond to Fig.~\ref{fig:three_evol1}. A cold cavity (random noise in each mode) initial conditions (red crosses) is used, the blue line represents the evolution, and the red asterisk represents the attained steady-state. In (c-d) we plot the solutions for $\delta=-4.5$, at fixed $P=40$, for two different initial conditions, cold and hot cavity. The former [dashed line, with green (light gray) markers] collapses rapidly to the lower branch fixed point, the latter (solid line) is attracted to an oscillating solution. }
\label{fig:phasespace3}
\end{figure}

The advantage of the present approach is to obtain the relative phases $\psi$ and $\phi_1$: in Fig.~\ref{fig:phasespace3}, we show the phase space evolution, in terms of polar representation in the two planes $(E\cos \phi_1,E\sin \phi_1)$ and $(\eta\cos \psi,\eta\sin \psi)$. (a-b) correspond to Fig.~\ref{fig:three_evol1}, (c-d) is the homologue of Fig.~5(c) in \cite{Armaroli2015cav} (with $\delta= -4.5$ and $P=40$, instead). 
In (a), it is apparent how the cold cavity is first excited into a state corresponding to an unstable non-oscillating solution, making a turn around it. Once $u_0$ gains enough energy [in (b), $\eta$ is attracted initially to 0, its initial value being random], it loses stability and converts abruptly ---about $\eta=0.22$ of its energy---to the $u_{\pm 1}$. The phases suddenly lock, to the values shown in the inset of Fig.~\ref{fig:bifurcation3}. In panel \ref{fig:phasespace3}(b), the random initial condition passes through $\eta=0$, then abruptly switches to a finite value with a locked $\phi_1$.

Panels (c-d) show the two possible scenarios, the initial condition being a cold (small random intensity) or a hot cavity (the central mode is already strongly excited, as shown by a red cross, $E(0)=2$).
The former, solid lines, just stabilizes to the fixed point lying on the lower branch (c), the rate of conversion drops to zero (d) (the apparent large initial $\eta$ is again an artifact of small random amplitudes). The latter is first attracted to the upper equilibrium branch, which is in turn unstable but allows the mode to enter in the oscillating regime. As it was shown in  \citep{Armaroli2015cav}, the two conditions are connected in the bifurcation diagram and we can switch adiabatically from one to another, by changing $\delta$. 
 
The region below $\delta=-4.5$ is quite richer, because a stable and unstable branches of limit cycles coexist with a stable and an unstable equilibria. The observation of these limit cycles is harder and harder, because, as we approach the limit point where the two cycles merge at $\delta= -6.65$, their basin of attraction shrinks. Moreover a bifurcation to tori exists, see the dotted interval in Fig.~\ref{fig:bifurcation3}.
 
\begin{figure}[hbtp]
\centering
\includegraphics[width=.45\textwidth]{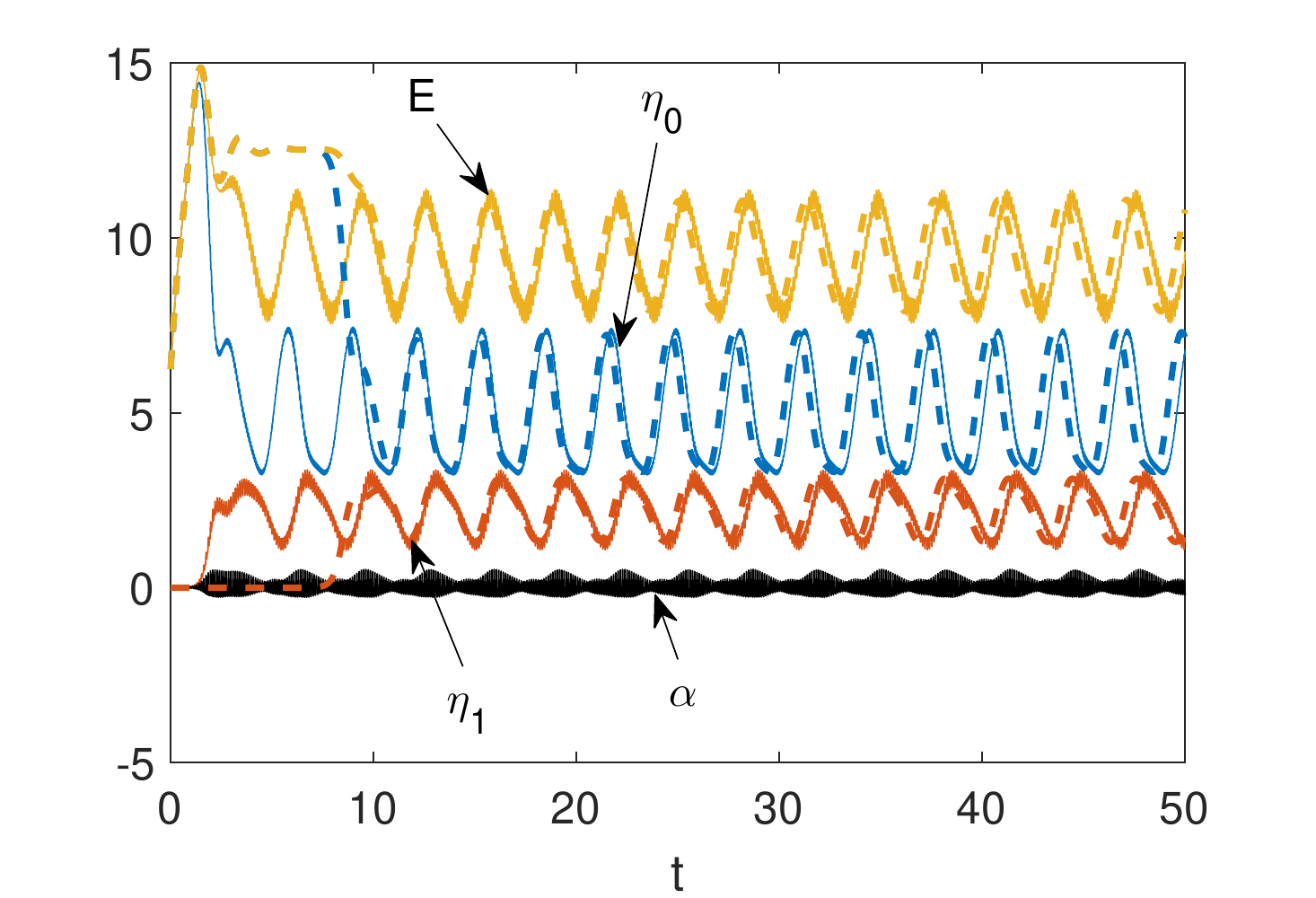}
\caption{(Color Online) Same as Fig.~\ref{fig:three_evol1}, but with  $\delta=-5.5$ and hot cavity initial conditions $u_0=2.5$ and $u_{\pm 1} = 0 $. The limit cycles are here unstable and the system oscillates on a torus, the secondary frequency is much smaller the main one. The average conversion to sidebands is about $25\%$.}
\label{fig:three_evol2}
\end{figure}

For the sake of completeness, we picture  this bifurcation by a means of a numerical example: we let $\delta=-5.5$  and an excited central super-mode  $u_0=2.5$ 	($u_{\pm1}=10^{-7}$).
The system settles on a limit cycle (oscillating at $\sqrt{2}\gamma$) modulated at a persistent low frequency  $\omega'\approx 2$, see Fig.~\ref{fig:three_evol2}. We did not discuss  this  bifurcation before, albeit it was observable in the adiabatic transition in Fig.~5(d) of \cite{Armaroli2015cav}, because it represents a source of noise and instability in view of the generation of microwaves on the optical carrier and should thus be avoided; however, this concerns a small region of the parameter space (see the dotted lines in Fig.~\ref{fig:bifurcation3}). The modal and super-modal approaches  agree quite well even in the present case, except for a slightly shorter period in the secondary oscillation predicted by the latter. $\alpha$ does not significantly deviate from $0$ over all the considered temporal range. 

\section{Four cavities}
\label{sec:4cav}

\subsection{Derivation of coupled-super-modes theory}

We now consider the $M=4$ system, depicted in Fig.~\ref{fig:scheme}(b). The most general symmetric system is $\gamma_{12}=\gamma_{34}=\gamma$ and $\gamma_{23}=\kappa\gamma$.

The  super-modes are located at frequencies $\Omega_{\pm 1}=\pm\Omega$ and $\Omega_{\pm 3}=\pm\Omega'$, with $\Omega \equiv \frac{1}{2}  \gamma  \left(\sqrt{\kappa ^2+4}- \kappa \right)$ and $\Omega' \equiv \frac{1}{2}  \gamma  \left(\sqrt{\kappa ^2+4}+ \kappa \right)$. We focus here on the specific case $\Omega'=3\Omega$, which is the most symmetric and efficient, as far as conversion is concerned, \cite{Trillo1994, MarhicBook2007, Ott2013c}. This condition is satisfied for $\kappa=\sqrt{4/3}$.
and allows us to obtain a simple eigenvector matrix
\[
	\mathbf{V}=\begin{bmatrix}
		q &q & p & p\\  -p &p & -q &q\\
		 -p& -p& q& q\\ q&-q&-p&p
	\end{bmatrix}
\]
with $p = \frac{\sqrt{2}}{4}$, $q = \frac{\sqrt{6}}{4}$.
We suppose the central modes are excited at the same time by a modulated input $f(t)=\cos{\Omega t}$ and study the energy conversion to $3\Omega$.

We now derive, along the same lines  of the previous section, the complex ODEs that govern the amplitudes $\mathbf{u}=[u_{-1},u_{1},u_{-3},u_{3}]^T$ (the subscripts referring to oscillations at multiples of frequency $\Omega$) of the super-modes--- {all oscillating terms are neglected,}
\begin{equation}
\begin{aligned}
	\dot{u}_{-1} &= (i\delta-1) {u}_{-1} + \bar \pi  \\
	&	+\frac{i\chi}{16} \left[\underbrace{\sqrt{3} u_{3}^* u_{1}^2 
	 +6 u_{-3} u_{3} u_{1}^* +  2 \sqrt{3} u_{-3} u_{-1}^* u_{1}}_\mathrm{Coh.}+
	   \right. \\ 
	  &\left.	   	
	   +\left( \underbrace{5 |u_{-1}|^2}_\mathrm{SPM} + \underbrace{10  |u_{1}|^2 + 6|u_{-3}|^2 +6 |u_{3}|^2}_\mathrm{XPM}\right)u_{-1}\right]
	\\
	\dot{u}_{1} &= (i\delta-1) {u}_{1}  + \bar \pi \\
	&+ \frac{i\chi}{16} \left[\underbrace{\sqrt{3} u_{-3}^* u_{-1}^2+6 u_{-3} u_{3} u_{-1}^* + 2 \sqrt{3} u_{3} u_{-1} u_{1}^*}_\mathrm{Coh.} + \right. \\ 
	&\left.
	+ \left(\underbrace{5 |u_{1}|^2}_\mathrm{SPM} + \underbrace{10 |u_{-1}|^2 + 6|u_{-3}|^2 + 6 |u_{3}|^2}_\mathrm{XPM}\right) u_{1}   \right]
	\\
	\dot{u}_{-3} &= (i\delta-1) {u}_{-3} + 
	\frac{i  \chi}{16} \left[\underbrace{ 6 u_{3}^* u_{-1} u_{1}}_\mathrm{Coh.}+\underbrace{\sqrt{3} u_{-1}^2 u_{1}^*}_\mathrm{Forced} + \right. \\
	&\left. 
	+\left(\underbrace{5 |u_{-3}|^2}_\mathrm{SPM} + \underbrace{10 |u_{3}|^2+6|u_{-1}|^2+6 |u_{1}|^2}_\mathrm{XPM}\right)u_{-3} \right]\\
	\dot{u}_{3} &= (i\delta-1) {u}_{3} + 
	 \frac{i\chi}{16} \left[\underbrace{6 u_{-3}^* u_{-1} u_{1}}_\mathrm{Coh.} + 
	\underbrace{ \sqrt{3}  u_{1}^2 u_{-1}^*}_\mathrm{Forced}+
	\right.\\
	&\left.
	+\left(\underbrace{5 |u_{3}|^2}_\mathrm{SPM} + \underbrace{10|u_{-3}|^2 + 6  |u_{-1}|^2+6|u_{1}|^2}_\mathrm{XPM}\right) u_{3} \right],
\end{aligned}
\label{eq:CSMT4}
\end{equation}
with $\bar \pi\equiv \frac{\sqrt{6 P}}{8}$.

The SPM, XPM and coherent interactions among super-modes are similar to those found in Eq.~\eqref{eq:CSMT3}. Moreover, the terms $\sqrt{3}  u_{\pm 1}^2 u_{\mp 1}^*$  originate from the particular choice of the FWM process, $\Omega+\Omega-(-\Omega) \to  3\Omega$ and act as a forcing at $3\Omega$. Thus the vanishing $u_{\pm 3}$ solution  $\eta_1\left[1+\left(\delta + \frac{15}{16}\chi \eta_1\right)^2\right] = \frac{3P}{32} = \bar \pi^2$ (with $\eta_1\equiv |u_1|^2$) is only approximate, because as soon as some energy is stored in $u_{\pm 1}$, it is partially converted to the lateral ones. The saddle-node bifurcation points are anyway well predicted by $\frac{15}{16}\chi \eta_1^\pm = -\frac{2}{3}\delta \pm \frac{\sqrt{\delta^2-3}}{3}$, at least at small $P$. As in Sec.~\ref{sec:3cav}, multiple equilibria occur for $\delta<-\sqrt{3}$.
Self-pulsing is basically thresholdless, as described in \cite{Hansson2014d}. 

As explained above, for large $\gamma$   spurious bifurcations and chaotic regimes are largely suppressed and, if $\Omega\gg |\delta|$, the pump couples evenly to both pump super-modes. Thus the imbalance of each harmonic pair, $\beta\equiv|u_1|^2-|u_{-1}|^2$ and $\alpha\equiv|u_3|^2-|u_{-3}|^2$, can be thus safely assumed to be zero. We will discuss  deviations below.

 {As in Sec.~\ref{sec:3cav}, in Eq.~\eqref{eq:CSMT4} we neglect non-phase-matched non-linear terms as well as non-resonant oscillating forcing. The full forcing vector reads as $\bar \pi [1 + \exp(i 2 \Omega),1+ \exp(-i 2 \Omega),\frac{\sqrt{3}}{3}(\exp(i 2 \Omega) + \exp(i 4 \Omega)),\frac{\sqrt{3}}{3}(\exp(-i 2 \Omega) + \exp(-i 4 \Omega))]^\mathrm{T}$: we verify numerically that oscillating terms act independently on each component (not shown) and are accounted for the oscillations observed in Figs.~\ref{fig:fourP180evol1}, \ref{fig:fourP180evol2}, and \ref{fig:fourP180evol3}.}

We derive the simplest real form for the system \eqref{eq:CSMT4}, by assuming $u_1=u_{-1}=\sqrt{\eta_1}\exp \left(i\phi_1\right)$ and $u_3=u_{-3}=\sqrt{\eta_3}\exp{(i\phi_3)}$ and  defining the positively detuned half energy $E\equiv \eta_1 +\eta_3$, the sideband fraction $\eta\equiv\eta_3/E$ and the relative phase $\psi\equiv \phi_3-\phi_1$. Together with the relative injection to $\Omega$-mode phase $\phi_1$, these variables evolve according to the following system,
\begin{equation}
\begin{aligned}
	\dot\eta &= 
	\eta(1-\eta)\left[\frac{6\chi E}{8}\sin{2\psi}+\frac{\sqrt{3}\chi E}{8} \sqrt{\frac{1-\eta}{\eta}}\sin{\psi} \right.
	\\
	&\left.- \frac{2\bar \pi}{\sqrt{E(1-\eta)}}
	\cos{\phi_1} \right] \\
	\dot{E} & = -2E + 2\bar {P} \sqrt{E(1-\eta)}\cos{\phi_1}\\
	\dot{\psi} &= \frac{\bar \pi}{\sqrt{E(1-\eta)}}\sin \phi_1 +  \frac{3\chi E}{16}\left[
		2\eta-1 
		\right.\\
		&\left.	
		+ 2 (1-2\eta) \cos{2\psi} 	 			
		+\frac{\sqrt{3\eta(1-\eta)} }{3}\left(\frac{1-\eta}{\eta}-3\right)\cos{\psi}	
	\right]\\
	\dot\phi_1 & = \delta - \frac{\bar \pi}{\sqrt{E(1-\eta)}}\sin \phi_1 
	\\	
	&+ \frac{3\chi E}{16}
	\left[	
		-\eta+5 +2\eta \cos{2\psi}+\sqrt{3\eta(1-\eta)}\cos\psi
	\right]
\end{aligned}
\label{eq:realCSMT4}
\end{equation}
For the sake of keeping the notation at a minimum, we  use the variable names of the previous section. Whenever we need to distinguish between them, we will add a superscript $M=3,4$.
It is straightforward to verify that $\eta=0,1$ do not correspond to admissible fixed points of Eq.~\eqref{eq:realCSMT4}, thus we study numerically the  other equilibria, which correspond necessarily to a small generated signal at $\pm 3 \Omega$.

\subsection{Bifurcation diagram and numerical results}

\begin{figure}[hbtp]
\centering
\includegraphics[width=0.45\textwidth]{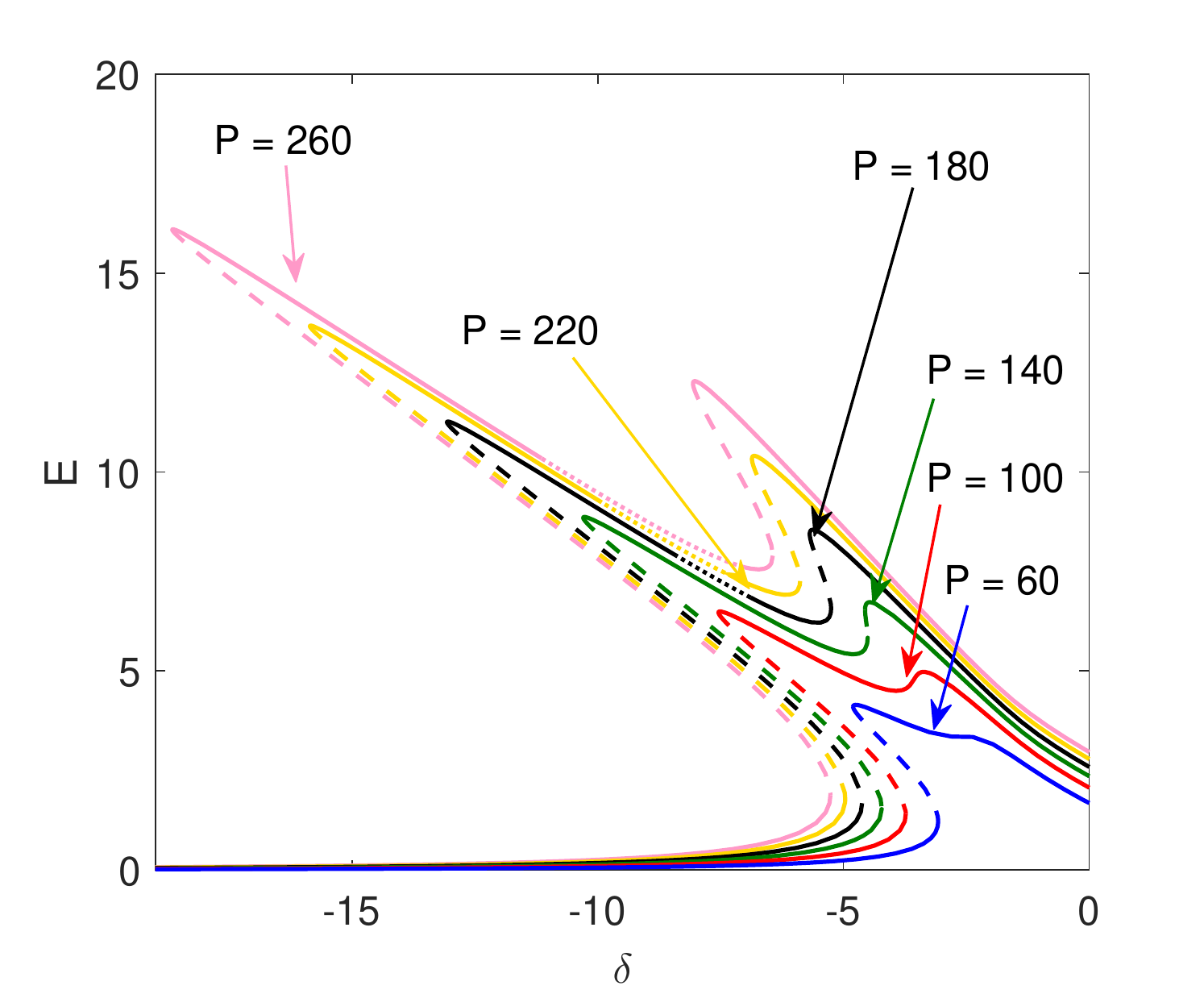}
\caption{(Color Online) Bifurcation diagram of $E$ ($\eta\neq 0,1$) as function of $\delta$ for different values of $P=60,100,140,180,220,260$. Solid  lines represent stable  limit cycles, dashed lines unstable (saddle points) and dotted lines are delimited by Neimark-Sacker  bifurcations. Each curve is labelled according to its value of $P$. The absence of an intensity-dependent threshold makes it impossible to find precise values for triggering the oscillations. Instead, as before in Fig.~\ref{fig:bifurcation3}, we have only a single branch of limit cycles for small $P<120$, while a saddle-node bifurcation is observed at higher $P>120$. An even more complex behavior is observed for larger injection power, but the super-mode approach is less accurate.}
\label{fig:4bifurcationE}
\end{figure}

\begin{figure}[hbtp]
\centering
\includegraphics[width=0.45\textwidth]{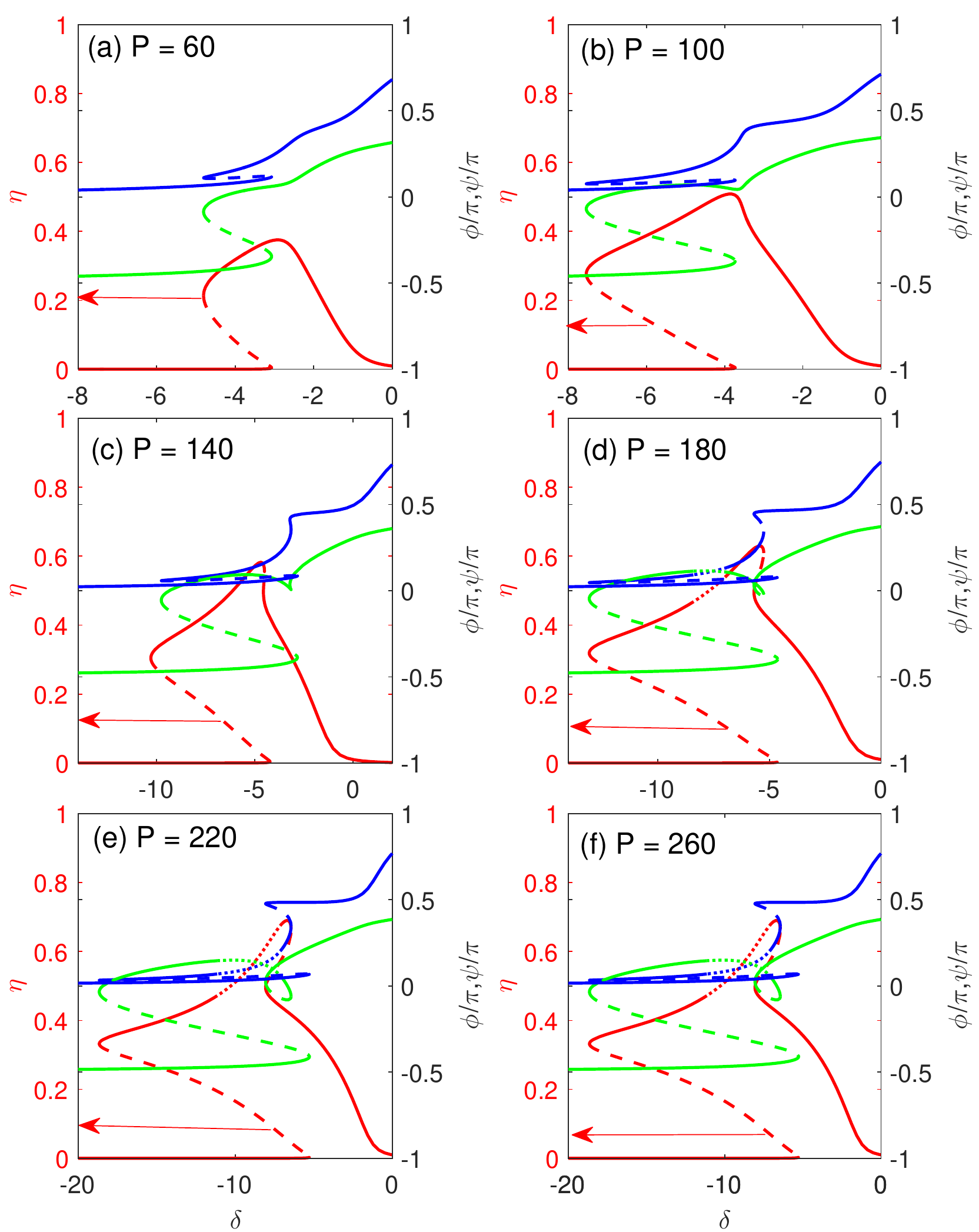}
\caption{(Color Online) Bifurcation diagrams of $\eta$ [red (gray), left ordinates, see arrow], $\psi$ [blue (dark grey), right ordinates] and $\phi$ [green (light gray), right ordinates]  for the  values of $P$ of Fig.~\ref{fig:4bifurcationE}. Notice that $\eta>0.5$, i.e.~more energy in lateral than in central modes.}
\label{fig:4bifurcationetaphase}
\end{figure}

In Fig.~\ref{fig:4bifurcationE} we show the bifurcation diagram of $E$ obtained from Eq.~\eqref{eq:realCSMT4} as a function of $\delta$, for different values of $P$ (the actual value of $\gamma\gg 1$ is not important, as it does not appear in Eq.~\eqref{eq:CSMT4}). Do not forget that $\delta$ is the detuning of the laser frequency (the center of the two pumps) with respect to the uncoupled linear cavity resonance, i.e.~the midpoint between $\pm\Omega$. Far from resonance, at $\Omega\gg |\delta|\gg 0$, the line connects smoothly with the undepleted pump solutions ($|u_{\pm 3}|\ll |u_{\pm 1}|$). This is the precise meaning of thresholdless excitation of the lateral resonances: contrary to the previous section, we do not need a specific combination of detuning and power levels in order to observe self-pulsing. As in the previous section,  multistability of limit cycles (two stable and one unstable branches coexist) occurs for $P>17.5$, as can be approximately predicted from $\eta\approx 0$. By increasing $P$, we observe that a second hump appears (at $P=60$, blue lower line, is already noticeable), then the hump folds towards negative $\delta$ and the branch of larger $E$ splits in  two stable and one unstable limit cycles. We now have three stable and two unstable branches: we will denote the stable limit cycles as upper, intermediate (the one extending towards extreme $\delta<0$), and lower.
The saddle-node bifurcation, where the lower stable and unstable branches merge is well approximated by the solution in the $\eta=0$ limit.
 The intermediate and upper stable solutions split farther and farther apart as we increase $P$. 
In Fig.~\ref{fig:4bifurcationetaphase} we show the bifurcation of $\eta$, $\phi_1$, and $\psi$. Notice that, on the intermediate branch, the conversion $\eta>0.5$, i.e.~more energy is trasnferred to $\pm 3\Omega$ than that in $\pm \Omega$.  In contrast to frequency combs \cite{Hansson2014d}, where the cascaded FWM distributes energy to higher and higher frequencies, depending on dispersion, and the spectral shaping demands complicated contrivances (feedback control of injection, dispersion engineering...), here an efficient energy conversion occurs spontaneously.
The upper branch is characterized by $\psi\approx\pi/2$, while the intermediate and lower ones require $\psi\approx 0$. 
The bifurcation analysis of Eq.~\eqref{eq:realCSMT4} predicts also a supercritical NS bifurcation from cycles to tori (for $P\gtrsim 80$). No other separate branches were found.

We present below some numerical examples of evolution in time in order to validate the super-modal approach and to understand the different permitted behaviors. We let $\gamma=40$, $\kappa = \sqrt{4/3}$, $P=180$ (as in Fig.~\ref{fig:4bifurcationetaphase}(d)).

\begin{figure}[hbtp]
\centering
\includegraphics[width=.45\textwidth]{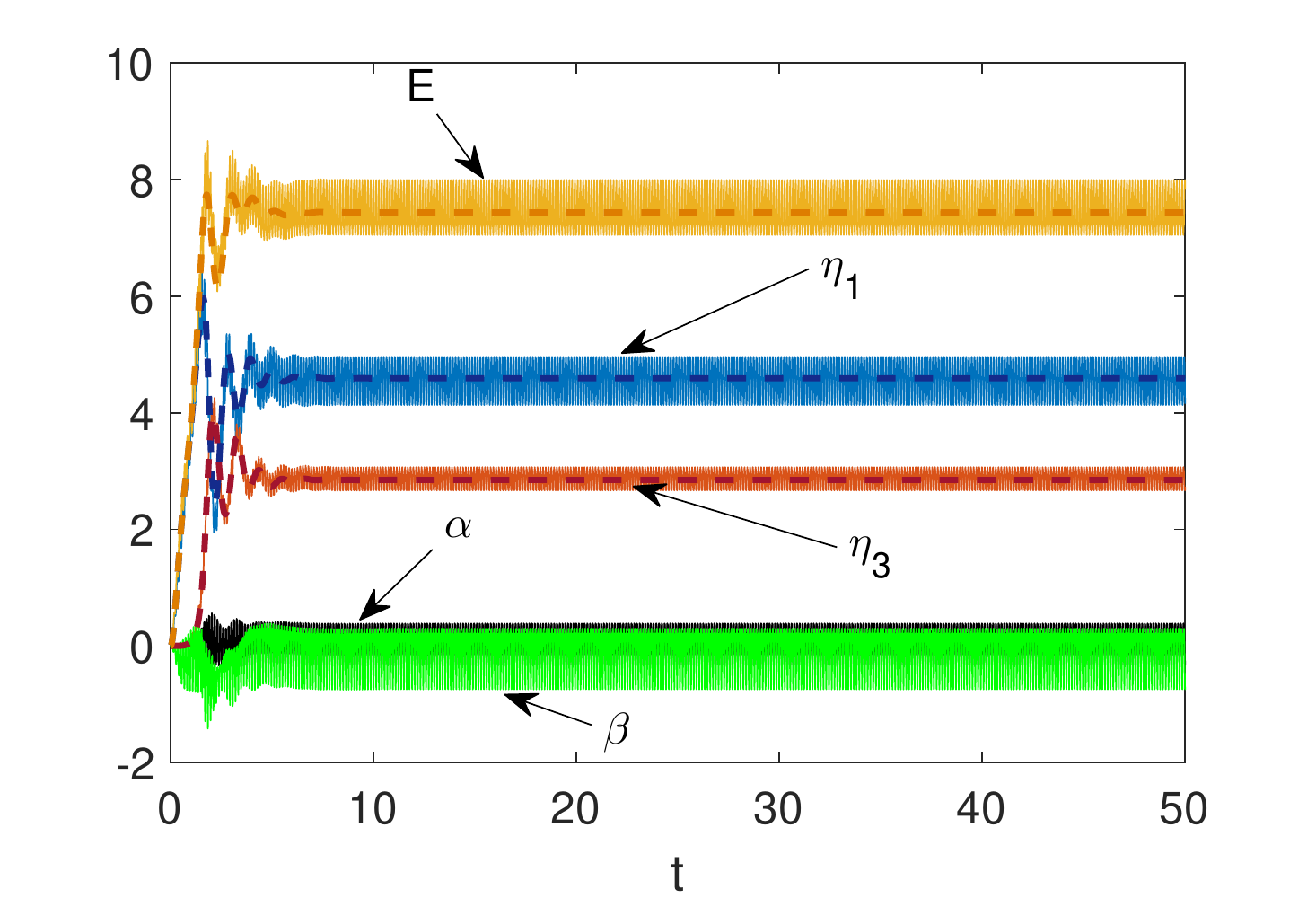}
\caption{(Color Online) Comparison of the evolution toward steady-state of the four cavity system and its four super-mode truncation. $P=180$, $\delta=-4.5$, $\gamma=40$. Solid lines represent the solution of Eq.~\eqref{eq:adim} ($M=4$), while dashed lines the solution of Eq.~\eqref{eq:CSMT4}: $\eta_1$ is in blue (dark gray), $\eta_3$ in red (gray), $E$ in yellow (light gray), the imbalances $\alpha$ ($\beta$) are finally at the bottom in black [green (light gray)] (only for Eq.~\eqref{eq:adim}). Notice that these last exhibit small oscillations around zero. }
\label{fig:fourP180evol1}
\end{figure}
Fig.~\ref{fig:fourP180evol1} shows the evolution towards a stable limit cycle, for $\delta=-4.5$, i.e.~where only the upper branch exists, starting from random noise-like initial perturbations (cold cavity). Notice that the pump and signal imbalances, $\beta$ and $\alpha$ oscillate around 0 during the whole time interval. A steady-state is quickly attained,  {apart from small oscillations at the non-resonantly forced $\Omega''=2\Omega$, see above}. The comparison of the numerical integration of Eq.~\eqref{eq:CSMT4} and of the conventional CMT, Eq.~\eqref{eq:adim}, shows a much better agreement than in the previous section, for $M=3$. This is due to the threshold-less nature of non-degenerate FWM, which provides an active forcing for $\pm 3 \Omega$. The phase-space representation [defined, as above, by the two planes $(E\cos\phi_1,E\sin\phi_1)$ and $(\eta\cos\psi,\eta\sin\psi)$] is provided in Fig.~\ref{fig:phasespace4}(a-b). They show how the energy starts soon converting to the side-modes, and does not need to heat the cavity before starting oscillations. This manifests itself in the absence of sharp phase jump away from a fixed point in (a) and the smooth growth of $\eta$ in (b); compare the present behavior to that of Fig.~\ref{fig:phasespace3}.

\begin{figure}[hbtp]
\centering
\includegraphics[width=.46\textwidth]{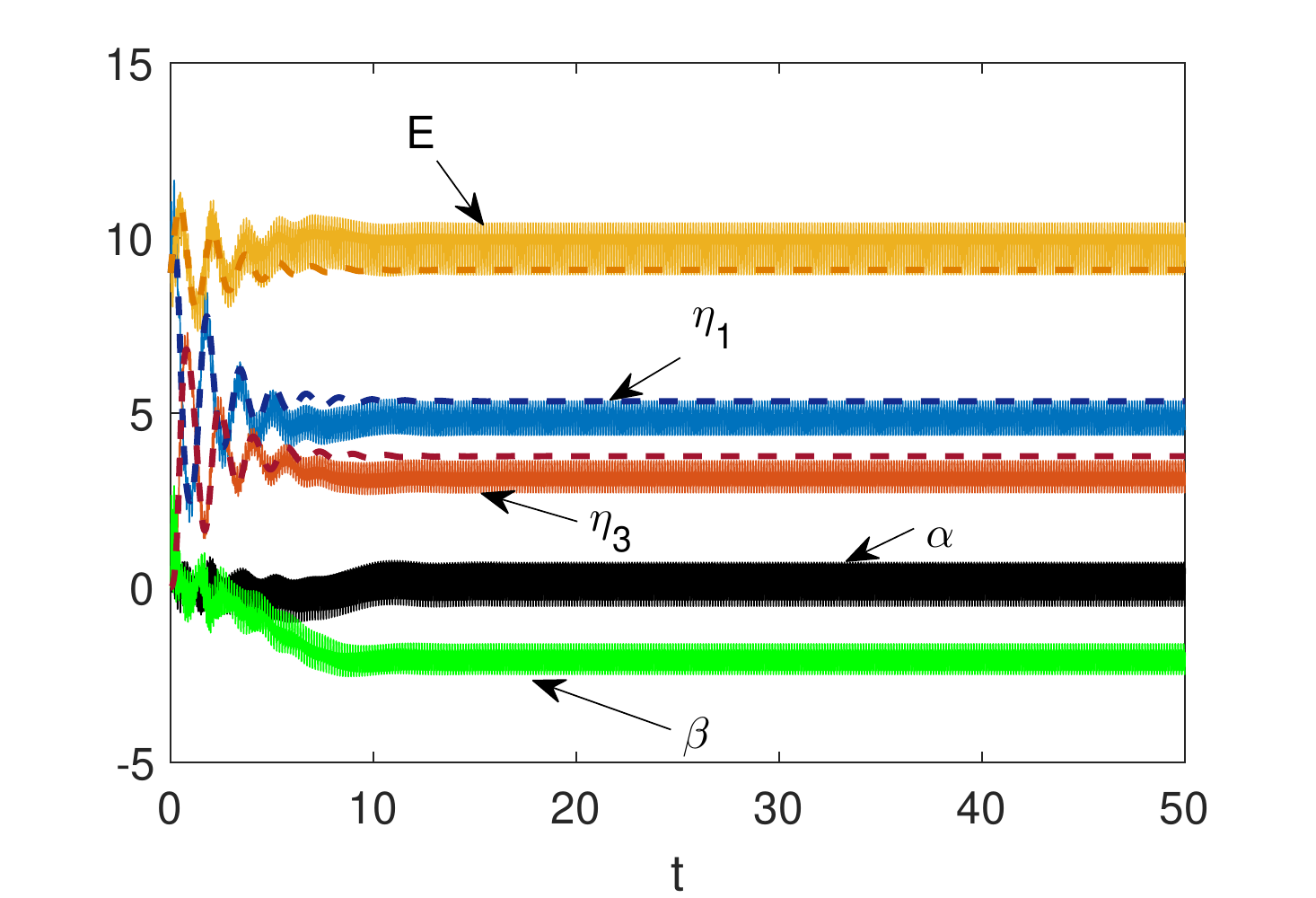}
\caption{(Color Online) Same as Fig.~\ref{fig:fourP180evol1}, for $\delta=-10$, and hot cavity initial conditions.}
\label{fig:fourP180evol2}
\end{figure}

Next, we study the more complex case, where the lower branch (where $\eta\approx 0$) coexists with the most energetic intermediate one. Let $\delta=-10$. Obviously, starting from a cold cavity leads the system to decay to the lower branch. We thus impose the hot cavity conditions $u_1=u_{-1}=3$ and $u_3=u_{-3}=0.001$, i.e.~the pump pair is already intense enough inside the cavity, so that it lies in the basin of attraction of the intermediate branch. The agreement between the two models is still satisfactory, although  $-\Omega$ is excited more strongly than its mirror $\Omega$---the imbalance $\beta$ turns indeed negative, while $\alpha$ oscillate steadily around 0. $\eta_{1,3}$  are only slightly overestimated, while $E$ is a bit underestimated. In Fig.~\ref{fig:phasespace4}(c-d) we map this solution in the pair of phase-planes. Notice that the initial conditions are very close to the final $E$ state, the basin of attraction of which is quite narrow (verified 
numerically, not shown). The system spirals around its steady-state. Panel (c) is quite similar to the dashed line in Fig.~\ref{fig:phasespace3}(c), while (d) differs owing to the thresholdless conversion mechanism: $\eta$ starts to grow at the very beginning.

\begin{figure}[hbtp]
\centering
\includegraphics[width=.46\textwidth]{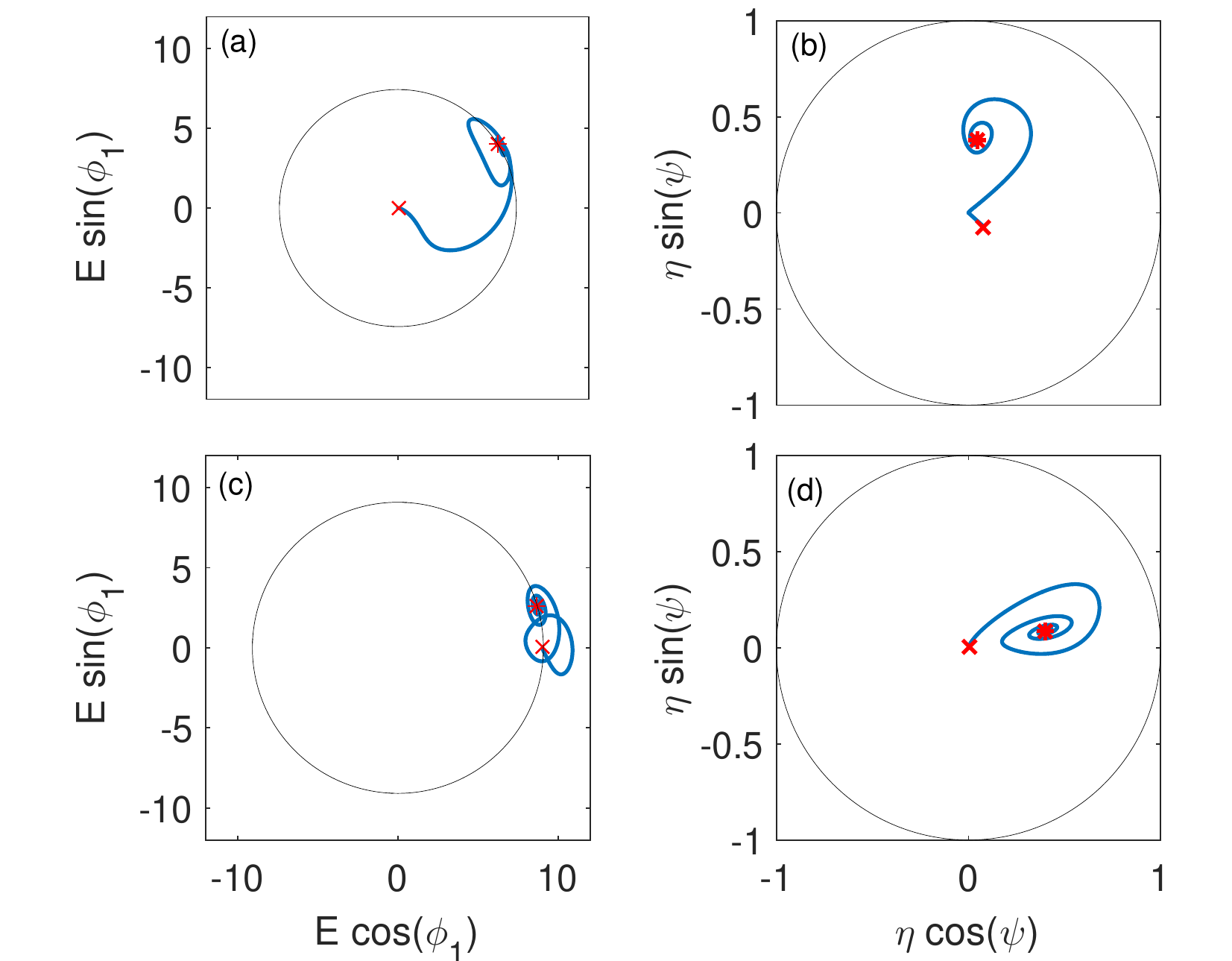}
\caption{(Color Online) Phase space representations of the time evolution of Eq.~\eqref{eq:CSMT4}. (a) and (c) show the plane $(E\cos\phi_1, E\sin\phi_1)$, (b) and (d) the plane $(\eta\cos\psi, \eta\sin\psi)$.  (a-b) Correspond to Fig.~\ref{fig:fourP180evol1}, (c-d) to Fig.~\ref{fig:fourP180evol2}, with hot cavity initial conditions. The notation is the same as in Fig.~\ref{fig:phasespace3}, but we omit to show the evolution from a cold cavity.}
\label{fig:phasespace4}
\end{figure}

\begin{figure}[hbtp]
\centering
\includegraphics[width=.45\textwidth]{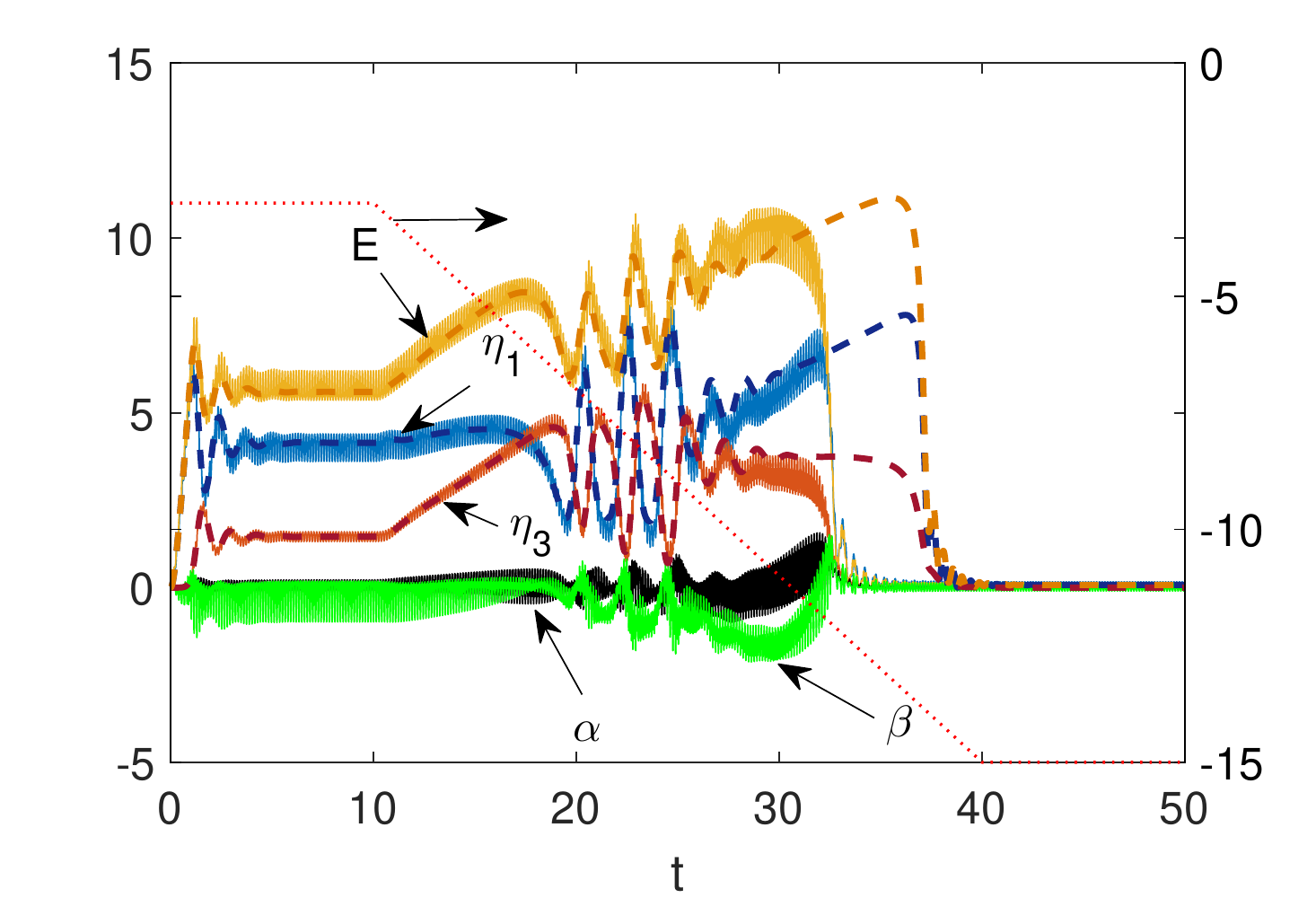}
\caption{(Color Online) Same as Fig.~\ref{fig:fourP180evol1}, for an adiabatic variation in $10<t<40$ of $\delta$ from -3 to -15. The right ordinate axis shows the values of $\delta$, plotted as a red (gray) dotted line. }
\label{fig:fourP180evol3}
\end{figure}
Finally we verify that two branches can be connected by an adiabatic variation of detuning (soft-excitability). We start from a cold cavity system and $\delta=-3$, let it stabilize up to $t=10$, then adiabatically decrease it up to $\delta=-15$ at $t=40$. In Fig.~\ref{fig:fourP180evol3} we show the results of our simulation. First we notice that the predicted existence and stability of the intermediate branch ends at $\delta=-12$ for Eq.~\eqref{eq:adim}, while at $\delta=-13$ for Eq.~\eqref{eq:CSMT4}, as predicted by the bifurcation diagram of Fig.~\ref{fig:4bifurcationE}.  This represents the main limitation of the present approximation. Then we observe that after the upper branch disappears at $\delta=-5.6$, strong oscillations set in, similar to Fig.~\ref{fig:three_evol2}. This confirms  the presence of the NS bifurcation shown in Figs.~\ref{fig:4bifurcationE} and \ref{fig:4bifurcationetaphase}.

For larger $P$ (not shown), it becomes harder to observe the most detuned part of the intermediate branch. The saddle-node bifurcation where the upper branch disappears occurs at nearly the same $\delta$ of the NS bifurcation. Moreover, larger region of existence for tori between the two NS points imply larger secondary oscillations. The  system is preferably attracted to the small conversion branch.

\section{Concrete implementation and limits}
\label{sec:implementation}

\subsection{Estimate of design parameters}

We finally comment on the physical accessibility of this approach. We assume to operate at $\lambda=1.55\,\mu\mathrm{\!m}$ and take $\tilde\tau_j=\tilde\tau=1\,$ns, so that the cavity has a {$Q=\frac{\tilde\omega\tilde\tau}{2}\approx 6.1 \times 10^5$}. Consider a racetrack cavity with minimum curvature radius $R=10\,\mu\mathrm{\!m}$ and mode area $A_\mathrm{eff}=1\,\mu{\rm\!m^2}$ (upper bound): the modal volume is $\mathcal{V} \approx 63\,\mu\mathrm{m^3}$.
and the effective index of the mode is $n_\mathrm{eff}=2$. A semiconductor of refractive index $3.48$ and Kerr index $n_2=2\times 10^{-17}\,\mathrm{m^2/W}$ is considered \cite{Wagner2009}, thus $\tilde\chi=2.90\times 10^{22}\,\mathrm{[J s]^{-1}}$; we finally get $I_0 =34.4\,$fJ.
We also assume weak waveguide coupling $\tilde\tau_\mathrm{wg}=10\tilde\tau$, so that the first cavity is undercoupled to the waveguide and the quality factor does not vary considerably from a cavity to the other.  

With these values, for $M=3$, $P=40$ corresponds to a power in the waveguide $|s_{in}|^2=6.90\,$mW {These power levels are feasible despite the undercoupling regime.}
As far as the coupling is concerned, a basic modal calculation, \cite{Little1997,HausBook}, permits to estimate that two waveguides of cross section $400\times300\,$nm with a gap of $200\,$nm require only about a coupling length $L_\mathrm{cpl}=4\,\mu\mathrm{\!m}$ to achieve the normalized value of $\gamma=40$.
This value was chosen in order to obtain an oscillation frequency  of about 9 GHz. The secondary frequency, where cycles bifurcate to tori is around 318 MHz.

The use of the same parameters for $M=4$ leads to a forcing (generated) frequency $\omega/2\pi = 3.67$GHz ($3\omega/2\pi = 11.0$ GHz). $P=180$ corresponds, for the considered sinusoidal forcing, to an average input power $\langle|s_\mathrm{in}|^2\rangle = 15.5$ mW. Still these values are attainable in current technological  platforms.

\subsection{Robustness to fabrication tolerances}

An important question is whether the fabrication tolerances with respect to nominal values inhibits the observation of the relevant phenomena.

We perform some Monte-Carlo simulations, by letting $\tilde\tau_j$, $\tilde\tau_\mathrm{wg}$, and $\tilde\gamma_{jk}$ randomly vary according to a Gaussian distribution around their nominal values; $P$ is kept constant. As in the previous paragraph the coupling time fluctuates around $\tau_\mathrm{wg}=10\tau$. We simulate $N_\mathrm{it}=2500$ different realizations of Eq.~\eqref{eq:adim} and look for the maximum standard deviation at which the system behaves as expected.

In the previous sections, we slightly abused of notation by using the same symbols for the autonomous $M=3$  and sinusoidally forced $M=4$ oscillators. Here we use superscripts to distinguish between them. 

For $M=3$ the system is very robust: an independent choice of every parameter with a standard deviation $\sigma=2.5\%$ is still tolerable. 
\begin{figure}[hbtp]
\centering
\includegraphics[width=.45\textwidth]{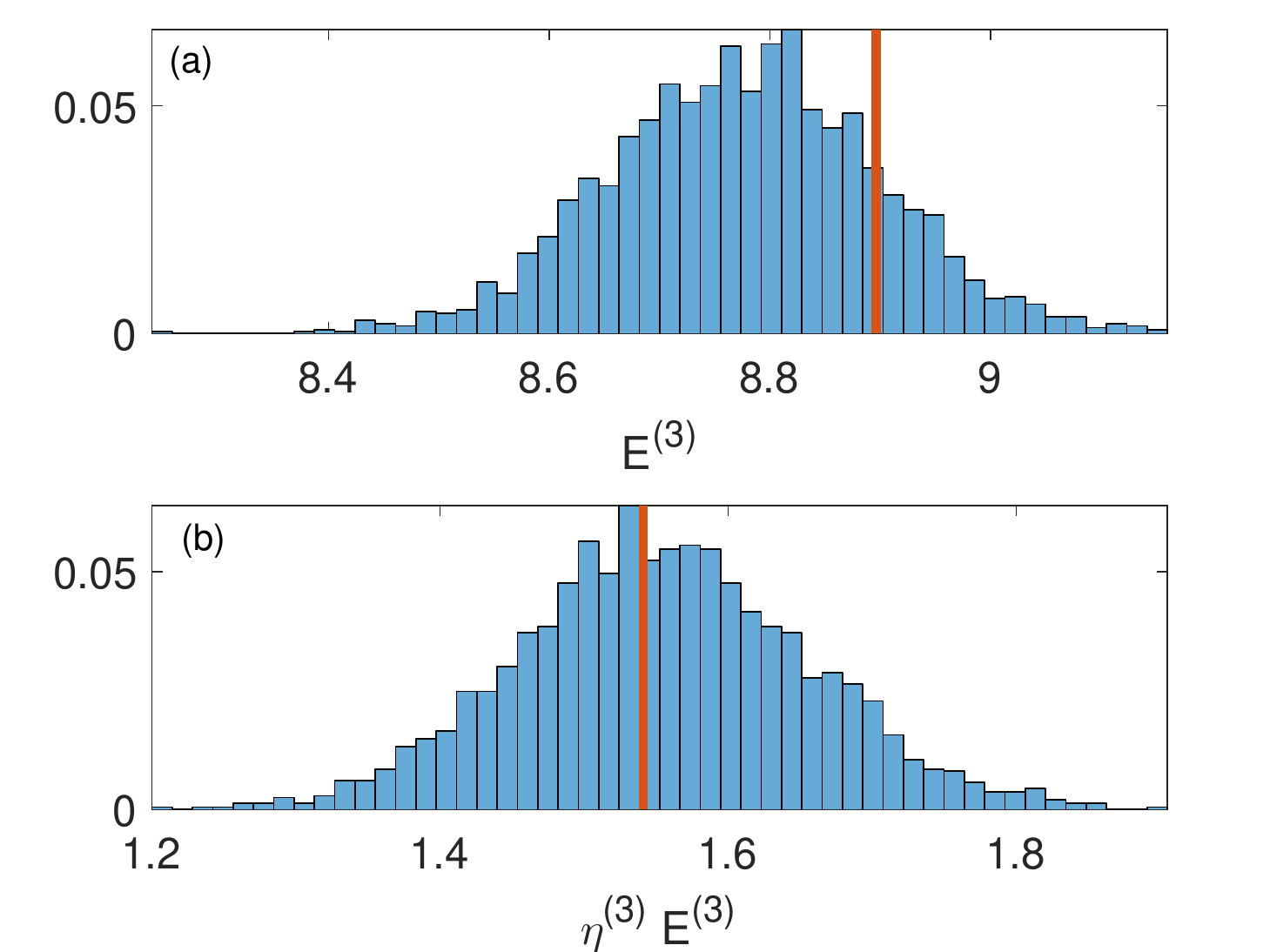}
\caption{(Color Online) Probability distribution of (a)  $E^{(3)}$ and (b) $\eta^{(3)} E^{(3)}$ obtained by repeatedly solving 2500 times the system of 
Eq.~\eqref{eq:adim} with $M=3$, $\delta=-3$, $\gamma=40$ and $P=40$. Every cavity parameter is normally distributed around their nominal value with a standard deviation of $2.5\%$. The thick red (gray) lines represent the nominal value.}
\label{fig:MC3}
\end{figure}
In Fig.~\ref{fig:MC3}, for $\delta= -3$ and $P=40$, the energy $E^{(3)}$  coupled in the cavity (a) and its fraction in the limit cycles $\eta^{(3)}E^{(3)}$ (b) are distributed around the predicted solution (see Fig.~\ref{fig:three_evol1}) with a standard deviation of about $3\%$ and $6\%$ respectively. The systematic inconsistency of the average $E^{(3)}$ (lower than expected) is explained by the inclusion of a finite external coupling $\tilde\tau_\mathrm{wg}$. The imbalance $\alpha^{(3)}$ (not shown) is distributed around 0 and does not represent a major problem. 

In the case of $M=4$, at each realization we assume to tune the input frequency in order to match the  eigenvalue $\pm\Omega$ of a pair of supermodes. This situation is more sensitive to small deviations, owing to the higher complexity of the system. Anyway a standard deviation $\sigma=1\%$ is well tolerated. 
\begin{figure}[ht]
\centering
\includegraphics[width=.45\textwidth]{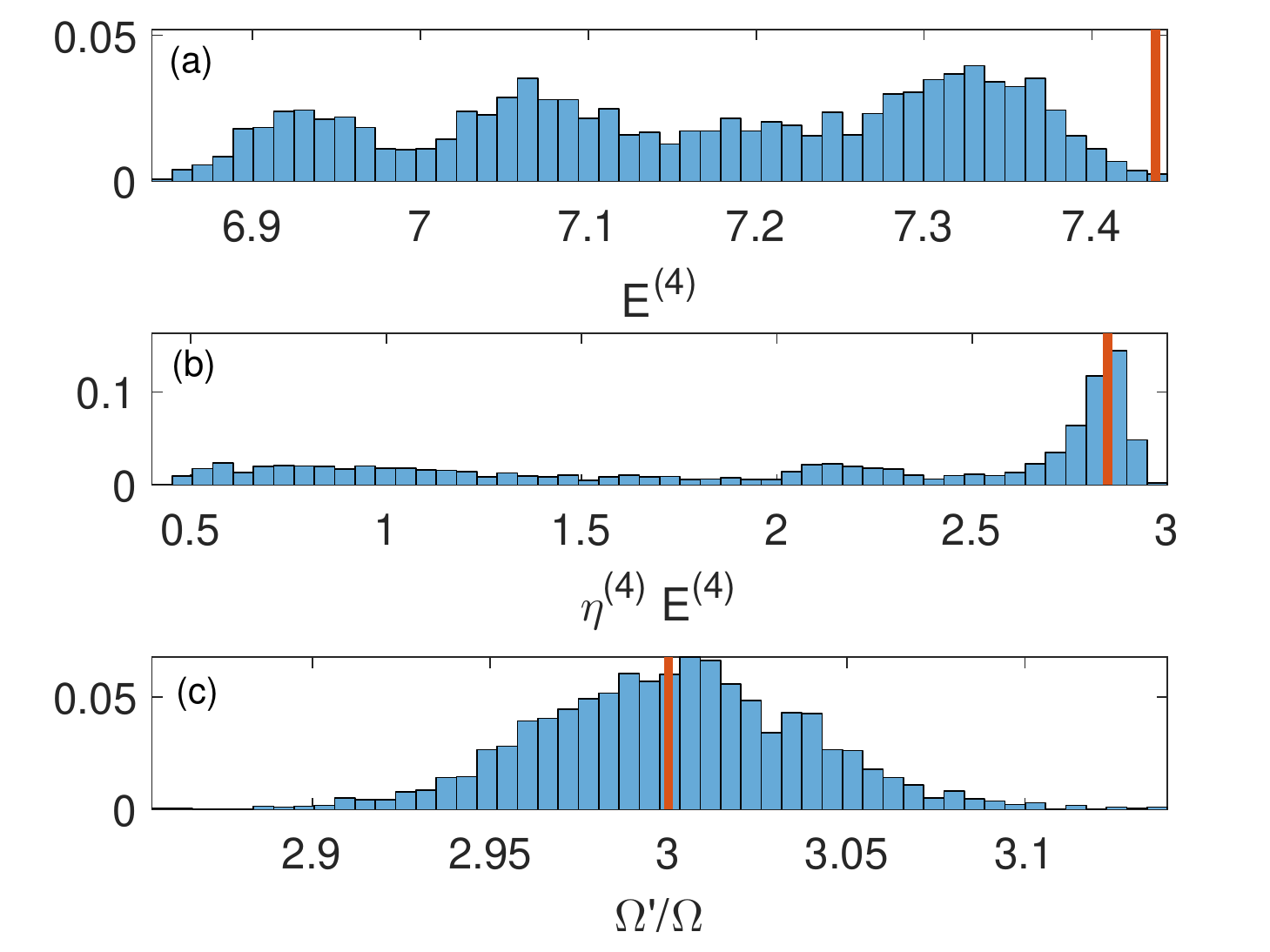}
\caption{(Color Online) Same as Fig.~\ref{fig:MC3}, for $M=4$ and a 1$\%$ variance for parameters around their nominal values ($\delta=-4$, $P=180$, $\gamma=40$, $\kappa=\sqrt{4/3}$). (a) $E^{(4)}$, (b) $\eta^{(4)}E^{(4)}$. (c) The distribution of the ratio of eigenfrequencies $\Omega'/\Omega$.}
\label{fig:MC4}
\end{figure}
This is shown in Fig.~\ref{fig:MC4}, for $\delta=-4.5$ and $P=180$: the energy stored in the cavity (a) is in most cases around the expected value (compare to Fig.~\ref{fig:fourP180evol1}), but with a difference limited to $10\%$ at most, on top of the systematic error mentioned above. The generated microwave energy (b) is peaked around the expected steady-state and exhibits a long tail of smaller values, too. Their distribution is much broader on account of the inefficient mechanism which occurs if $\Omega'/\Omega\neq 3$. This latter is shown in panel (c) to be distributed according to a Gaussian curve with a standard deviation of about $1\%$. The main limitation on conversion efficiency is thus  the spacing of the super-mode resonances compared to their lifetime, but FWM is effective in more than half of the cases. Reducing the uncertainty on parameters improves the results.  {The fine tuning techniques available today, see e.g.~\cite{Okawachi2015}, allow one to well adapt the system to the source frequency instead and compensate for fabrication imperfections; furthermore our approach is more compact than more conventional ones based on non-linear effects in non-resonant configurations \cite{Zheng2014,Long2016}.}

\section{Conclusions}
\label{sec:conclusion}

Based on a diagonalization procedure of the coupled-mode theory in time, which allows us to write the nonlinear equations which rule the coupling between super-modes in a chain of Kerr-nonlinear optical micro-resonators, we present a thorough bifurcation analysis of (i) degenerate four-wave mixing in a three cavity chain (ii) non-degenerate four-wave mixing in a four cavity chain. Their bifurcation diagrams and behavior in phase space are similar in many aspects (multistability of  limit cycles, NS bifurcations, phase locking between injection and pump and between pump and sidebands). The main difference relies on the thresholdless microwave generation in the four-cavity system. 

A sensible set of parameters is presented to show the accessibility of these oscillatory regimes. Moreover the exploration of the parameter space by means of Monte-Carlo simulations allows us to estimate the robustness of the present solutions to technological inaccuracies. The result indicates that the four cavity solution is less robust (tolerating quite a smaller uncertainty level), but still achievable in current technological platform.

Our results can be applied also to different architectures \cite{Combrie2017,Okawachi2015} and pave the way to the fabrication of microwave oscillators and converters on an optical carrier and the optimization of frequency combs in optical micro- and nano-cavities.  

\section*{Acknowledgments}
Y.~D. acknowledges the support of the Institut Universitaire de France (IUF).


\begin{thebibliography}{43}%
\makeatletter
\providecommand \@ifxundefined [1]{%
 \@ifx{#1\undefined}
}%
\providecommand \@ifnum [1]{%
 \ifnum #1\expandafter \@firstoftwo
 \else \expandafter \@secondoftwo
 \fi
}%
\providecommand \@ifx [1]{%
 \ifx #1\expandafter \@firstoftwo
 \else \expandafter \@secondoftwo
 \fi
}%
\providecommand \natexlab [1]{#1}%
\providecommand \enquote  [1]{``#1''}%
\providecommand \bibnamefont  [1]{#1}%
\providecommand \bibfnamefont [1]{#1}%
\providecommand \citenamefont [1]{#1}%
\providecommand \href@noop [0]{\@secondoftwo}%
\providecommand \href [0]{\begingroup \@sanitize@url \@href}%
\providecommand \@href[1]{\@@startlink{#1}\@@href}%
\providecommand \@@href[1]{\endgroup#1\@@endlink}%
\providecommand \@sanitize@url [0]{\catcode `\\12\catcode `\$12\catcode
  `\&12\catcode `\#12\catcode `\^12\catcode `\_12\catcode `\%12\relax}%
\providecommand \@@startlink[1]{}%
\providecommand \@@endlink[0]{}%
\providecommand \url  [0]{\begingroup\@sanitize@url \@url }%
\providecommand \@url [1]{\endgroup\@href {#1}{\urlprefix }}%
\providecommand \urlprefix  [0]{URL }%
\providecommand \Eprint [0]{\href }%
\providecommand \doibase [0]{http://dx.doi.org/}%
\providecommand \selectlanguage [0]{\@gobble}%
\providecommand \bibinfo  [0]{\@secondoftwo}%
\providecommand \bibfield  [0]{\@secondoftwo}%
\providecommand \translation [1]{[#1]}%
\providecommand \BibitemOpen [0]{}%
\providecommand \bibitemStop [0]{}%
\providecommand \bibitemNoStop [0]{.\EOS\space}%
\providecommand \EOS [0]{\spacefactor3000\relax}%
\providecommand \BibitemShut  [1]{\csname bibitem#1\endcsname}%
\let\auto@bib@innerbib\@empty
\bibitem [{\citenamefont {Ilchenko}\ and\ \citenamefont
  {Matsko}(2006)}]{Ilchenko2006}%
  \BibitemOpen
  \bibfield  {author} {\bibinfo {author} {\bibfnamefont {V.~S.~}
  \bibnamefont {Ilchenko}}\ and\ \bibinfo {author} {\bibfnamefont {A.~B. }\
  \bibnamefont {Matsko}},\ }\bibfield  {title} {\enquote {\bibinfo {title}
  {{Optical resonators with whispering-gallery modes—Part II:
  Applications}},}\ }\href
  {http://scholar.google.com/scholar?hl=en{\&}btnG=Search{\&}q=intitle:Optical+resonators+with+whispering-gallery+modes-Part+II:+applications{\#}0}
  {\bibfield  {journal} {\bibinfo  {journal} {IEEE J.~Sel.Top.~Quant.~Electron.~} }\textbf {\bibinfo {volume} {12}},\ \bibinfo {pages}
  {15--32} (\bibinfo {year} {2006})}\BibitemShut {NoStop}%
\bibitem [{\citenamefont {Turner}\ \emph {et~al.}(2008)\citenamefont {Turner},
  \citenamefont {Foster}, \citenamefont {Gaeta},\ and\ \citenamefont
  {Lipson}}]{Turner2008}%
  \BibitemOpen
  \bibfield  {author} {\bibinfo {author} {\bibfnamefont {A.~C.~} \bibnamefont
  {Turner}}, \bibinfo {author} {\bibfnamefont {M.~A.}\ \bibnamefont
  {Foster}}, \bibinfo {author} {\bibfnamefont {A.~L.~} \bibnamefont
  {Gaeta}}, \ and\ \bibinfo {author} {\bibfnamefont {M.~} \bibnamefont
  {Lipson}},\ }\bibfield  {title} {\enquote {\bibinfo {title} {{Ultra-low power
  parametric frequency conversion in a silicon microring resonator.}}}\ }\href
  {http://www.ncbi.nlm.nih.gov/pubmed/18542587} {\bibfield  {journal} {\bibinfo
   {journal} {Opt.~Express}\ }\textbf {\bibinfo {volume} {16}},\ \bibinfo
  {pages} {4881--4887} (\bibinfo {year} {2008})}\BibitemShut {NoStop}%
\bibitem [{\citenamefont {Azzini}\ \emph {et~al.}(2013)\citenamefont {Azzini},
  \citenamefont {Grassani}, \citenamefont {Galli}, \citenamefont {Gerace},
  \citenamefont {Patrini}, \citenamefont {Liscidini}, \citenamefont {Velha},\
  and\ \citenamefont {Bajoni}}]{Azzini2013b}%
  \BibitemOpen
  \bibfield  {author} {\bibinfo {author} {\bibfnamefont {S.~} \bibnamefont
  {Azzini}}, \bibinfo {author} {\bibfnamefont {D.~} \bibnamefont
  {Grassani}}, \bibinfo {author} {\bibfnamefont {M. ~} \bibnamefont
  {Galli}}, \bibinfo {author} {\bibfnamefont {D.~} \bibnamefont {Gerace}},
  \bibinfo {author} {\bibfnamefont {M.}\ \bibnamefont {Patrini}},
  \bibinfo {author} {\bibfnamefont {M.~} \bibnamefont {Liscidini}}, \bibinfo
  {author} {\bibfnamefont {P.~} \bibnamefont {Velha}}, \ and\ \bibinfo
  {author} {\bibfnamefont {D.~} \bibnamefont {Bajoni}},\ }\bibfield
  {title} {\enquote {\bibinfo {title} {{Stimulated and spontaneous four-wave
  mixing in silicon-on-insulator coupled photonic wire nano-cavities}},}\
  }\href {\doibase 10.1063/1.4812640} {\bibfield  {journal} {\bibinfo
  {journal} {Appl.~Phys.~Lett.}\ }\textbf {\bibinfo {volume} {103}},\
  \bibinfo {pages} {10--14} (\bibinfo {year} {2013})},\ 
  \BibitemShut {NoStop}%
\bibitem [{\citenamefont {Pu}\ \emph {et~al.}(2015)\citenamefont {Pu},
  \citenamefont {Hu}, \citenamefont {Ottaviano}, \citenamefont {Semenova},
  \citenamefont {Vukovic}, \citenamefont {Oxenlowe},\ and\ \citenamefont
  {Yvind}}]{Pu2015}%
  \BibitemOpen
  \bibfield  {author} {\bibinfo {author} {\bibfnamefont {M.~} \bibnamefont
  {Pu}}, \bibinfo {author} {\bibfnamefont {H.~} \bibnamefont{Hu}}, \bibinfo
  {author} {\bibfnamefont {L.~} \bibnamefont {Ottaviano}}, \bibinfo {author}
  {\bibfnamefont {E.~} \bibnamefont {Semenova}}, \bibinfo {author}
  {\bibfnamefont {D.~} \bibnamefont {Vukovic}}, \bibinfo {author}
  {\bibfnamefont {L.~K.~} \bibnamefont {Oxenlowe}}, \ and\ \bibinfo {author}
  {\bibfnamefont {K.~} \bibnamefont {Yvind}},\ }\bibfield  {title}
  {\enquote {\bibinfo {title} {{AlGaAs-On-Insulator Nanowire with 750 nm FWM
  Bandwidth, -9 dB CW Conversion Efficiency, and Ultrafast Operation Enabling
  Record Tbaud Wavelength Conversion}},}\ }in\ \href {\doibase
  10.1364/OFC.2015.Th5A.3} {\emph {\bibinfo {booktitle} {Optical Fiber
  Communication Conference Post Deadline Papers}}},\ \bibinfo {series and
  number} {\bibinfo {number} {September}}\ (\bibinfo {year} {2015})\ paper\
  \bibinfo {pages} {Th5A.3}\BibitemShut {NoStop}%
\bibitem [{\citenamefont {Combri{\'{e}}}\ \emph {et~al.}(2017)\citenamefont
  {Combri{\'{e}}}, \citenamefont {Martin},\ and\ \citenamefont
  {Rossi}}]{Combrie2017}%
  \BibitemOpen
  \bibfield  {author} {\bibinfo {author} {\bibfnamefont {S.~} \bibnamefont
  {Combri{\'{e}}}}, \bibinfo {author} {\bibfnamefont {A.~} \bibnamefont
  {Martin}}, \ and\ \bibinfo {author} {\bibfnamefont {A.~} \bibnamefont
  {de~Rossi}},\ }\bibfield  {title} {\enquote {\bibinfo {title} {{Comb of high-Q
  Resonances in a Compact Photonic Cavity}},}\ }{\bibinfo
  {journal} {Laser Photonics Rev.}\ }\href {\doibase
  10.1002/lpor.201700099} {\ \textbf {\bibinfo {volume} {11}},\ \bibinfo
  {pages} {1700099} (\bibinfo {year} {2017})}\BibitemShut {NoStop}%
\bibitem [{\citenamefont {Ibanescu}\ \emph {et~al.}(2002)\citenamefont
  {Ibanescu}, \citenamefont {Johnson}, \citenamefont {Fink}, \citenamefont
  {Joannopoulos}, \citenamefont {Soljac},\ and\ \citenamefont
  {Solja{\v{c}}i{\'{c}}}}]{Ibanescu2002}%
  \BibitemOpen
\bibinfo {author} {\bibfnamefont {M.~} \bibnamefont
  {Soljacic},  
  \bibfield  {author} {\bibfnamefont {M.~} \bibnamefont
  {Ibanescu}}, \bibinfo {author} {\bibfnamefont {S.~G.~} \bibnamefont
  {Johnson}}, \bibinfo {author} {\bibfnamefont {Y.~} \bibnamefont {Fink}}, and 
  \bibinfo {author} {\bibfnamefont {John~D}\ \bibnamefont {Joannopoulos}},
 \ }\bibfield  {title} {\enquote {\bibinfo {title}
  {{Optimal bistable switching in nonlinear photonic crystals}},}\ }\href
  {\doibase 10.1103/PhysRevE.66.055601} {\bibfield  {journal} {\bibinfo
  {journal} {Phys.~Rev.~E}\ }\textbf {\bibinfo {volume} {66}},\ \bibinfo
  {pages} {055601} (\bibinfo {year} {2002})}\BibitemShut {NoStop}%
\bibitem [{\citenamefont {Sylvain}\ \emph {et~al.}(2013)\citenamefont
  {Sylvain}, \citenamefont {Lehoucq}, \citenamefont {Junay}, \citenamefont
  {Malaguti}, \citenamefont {Bellanca}, \citenamefont {Trillo}, \citenamefont
  {Ménager}, \citenamefont {Reithmaier},\ and\ \citenamefont {{De
  Rossi}}}]{Combrie2013}%
  \BibitemOpen
  \bibfield  {author} {\bibinfo {author} {\bibfnamefont {S.~Combri{\'{e}}}}, \bibinfo {author} {\bibfnamefont {G.~}
  \bibnamefont {Lehoucq}}, \bibinfo {author} {\bibfnamefont {A.~}
  \bibnamefont {Junay}}, \bibinfo {author} {\bibfnamefont {S.~}
  \bibnamefont {Malaguti}}, \bibinfo {author} {\bibfnamefont {G.~}
  \bibnamefont {Bellanca}}, \bibinfo {author} {\bibfnamefont {S.~}
  \bibnamefont {Trillo}}, \bibinfo {author} {\bibfnamefont {L.~} \bibnamefont
  {Ménager}}, \bibinfo {author} {\bibfnamefont {J.~P.~} \bibnamefont
  {Reithmaier}}, \ and\ \bibinfo {author} {\bibfnamefont {A.~}
  \bibnamefont {{De Rossi}}},\ }\bibfield  {title} {\enquote {\bibinfo {title}
  {{All-optical signal processing at 10 GHz using a photonic crystal
  molecule}},}\ }\href {\doibase 10.1063/1.4829556} {\bibfield  {journal}
  {\bibinfo  {journal} {Appl.~Phys.~Lett.}\ }\textbf {\bibinfo {volume}
  {103}},\ \bibinfo {pages} {193510} (\bibinfo {year} {2013})}\BibitemShut
  {NoStop}%
\bibitem [{\citenamefont {Ghisa}\ \emph {et~al.}(2007)\citenamefont {Ghisa},
  \citenamefont {Dumeige}, \citenamefont {{Nguy{\^{e}}n Thi Kim}},
  \citenamefont {Boucher},\ and\ \citenamefont {Feron}}]{Ghisa2007}%
  \BibitemOpen
  \bibfield  {author} {\bibinfo {author} {\bibfnamefont {L.~} \bibnamefont
  {Ghisa}}, \bibinfo {author} {\bibfnamefont {Y.~} \bibnamefont
  {Dumeige}}, \bibinfo {author} {\bibfnamefont {N.~} \bibnamefont
  {{Nguy{\^{e}}n Thi Kim}}}, \bibinfo {author} {\bibfnamefont {Y.~G.~}
  \bibnamefont {Boucher}}, \ and\ \bibinfo {author} {\bibfnamefont {P.~}
  \bibnamefont {Feron}},\ }\bibfield  {title} {\enquote {\bibinfo {title}
  {{Performances of a fully integrated all-optical pulse reshaper based on
  cascaded coupled nonlinear microring resonators}},}\ }\href {\doibase
  10.1109/JLT.2007.901529} {\bibfield  {journal} {\bibinfo  {journal} {J.~Lightwave Tech.~} }\textbf {\bibinfo {volume} {25}},\ \bibinfo
  {pages} {2417--2426} (\bibinfo {year} {2007})}\BibitemShut {NoStop}%
\bibitem [{\citenamefont {Matsko}\ \emph {et~al.}(2005)\citenamefont {Matsko},
  \citenamefont {Savchenkov}, \citenamefont {Strekalov}, \citenamefont
  {Ilchenko},\ and\ \citenamefont {Maleki}}]{Matsko2005b}%
  \BibitemOpen
  \bibfield  {author} {\bibinfo {author} {\bibfnamefont {A.~B.~}
  \bibnamefont {Matsko}}, \bibinfo {author} {\bibfnamefont {A.~A.~}
  \bibnamefont {Savchenkov}}, \bibinfo {author} {\bibfnamefont {D.~}
  \bibnamefont {Strekalov}}, \bibinfo {author} {\bibfnamefont {V.~S.~}
  \bibnamefont {Ilchenko}}, \ and\ \bibinfo {author} {\bibfnamefont {L.~}
  \bibnamefont {Maleki}},\ }\bibfield  {title} {\enquote {\bibinfo {title}
  {{Optical hyperparametric oscillations in a whispering-gallery-mode
  resonator: Threshold and phase diffusion}},}\ }\href {\doibase
  10.1103/PhysRevA.71.033804} {\bibfield  {journal} {\bibinfo  {journal}
  {Phys.~Rev.~A }\ }\textbf
  {\bibinfo {volume} {71}},\ \bibinfo {pages} {033804} (\bibinfo {year}
  {2005})}\BibitemShut {NoStop}%
\bibitem [{\citenamefont {Del'Haye}\ \emph {et~al.}(2008)\citenamefont
  {Del'Haye}, \citenamefont {Arcizet}, \citenamefont {Schliesser},
  \citenamefont {Holzwarth},\ and\ \citenamefont {Kippenberg}}]{DelHaye2008}%
  \BibitemOpen
  \bibfield  {author} {\bibinfo {author} {\bibfnamefont {P.~}\bibnamefont
  {Del'Haye}}, \bibinfo {author} {\bibfnamefont {O.}~\bibnamefont {Arcizet}},
  \bibinfo {author} {\bibfnamefont {A.}~\bibnamefont {Schliesser}}, \bibinfo
  {author} {\bibfnamefont {R.}~\bibnamefont {Holzwarth}}, \ and\ \bibinfo
  {author} {\bibfnamefont {T.~J.~}\bibnamefont {Kippenberg}},\ }\bibfield
  {title} {\enquote {\bibinfo {title} {{Full stabilization of a
  microresonator-based optical frequency comb}},}\ }\href {\doibase
  10.1103/PhysRevLett.101.053903} {\bibfield  {journal} {\bibinfo  {journal}
  {Phys.~Rev.~Lett.}\ }\textbf {\bibinfo {volume} {101}},\ \bibinfo
  {pages} {053903} (\bibinfo {year} {2008})},\  \BibitemShut {NoStop}%
\bibitem [{\citenamefont {Razzari}\ \emph {et~al.}(2010)\citenamefont
  {Razzari}, \citenamefont {Duchesne}, \citenamefont {Ferrera}, \citenamefont
  {Morandotti}, \citenamefont {Chu}, \citenamefont {Little},\ and\
  \citenamefont {Moss}}]{Duchesne2010}%
  \BibitemOpen
  \bibfield  {author} {\bibinfo {author} {\bibfnamefont {L.}~\bibnamefont
  {Razzari}}, \bibinfo {author} {\bibfnamefont {D.}~\bibnamefont {Duchesne}},
  \bibinfo {author} {\bibfnamefont {M.}~\bibnamefont {Ferrera}}, \bibinfo
  {author} {\bibfnamefont {R.}~\bibnamefont {Morandotti}}, \bibinfo {author}
  {\bibfnamefont {S.}~\bibnamefont {Chu}}, \bibinfo {author} {\bibfnamefont
  {B.~E.}\ \bibnamefont {Little}}, \ and\ \bibinfo {author} {\bibfnamefont
  {D.~J.}\ \bibnamefont {Moss}},\ }\bibfield  {title} {\enquote {\bibinfo
  {title} {{CMOS-compatible integrated optical hyper-parametric oscillator}},}\
  }\href {\doibase 10.1038/nphoton.2009.236} {\bibfield  {journal} {\bibinfo
  {journal} {Nat.~Photonics}\ }\textbf {\bibinfo {volume} {4}},\ \bibinfo
  {pages} {41--45} (\bibinfo {year} {2010})}\BibitemShut {NoStop}%
\bibitem [{\citenamefont {Del'Haye}\ \emph {et~al.}(2009)\citenamefont
  {Del'Haye}, \citenamefont {Arcizet}, \citenamefont {Gorodetsky},
  \citenamefont {Holzwarth},\ and\ \citenamefont {Kippenberg}}]{Del'Haye2009}%
  \BibitemOpen
  \bibfield  {author} {\bibinfo {author} {\bibfnamefont {P.~} \bibnamefont
  {Del'Haye}}, \bibinfo {author} {\bibfnamefont {O.~}\bibnamefont {Arcizet}},
  \bibinfo {author} {\bibfnamefont {M.~L.~} \bibnamefont {Gorodetsky}},
  \bibinfo {author} {\bibfnamefont {R.}~\bibnamefont {Holzwarth}}, \ and\
  \bibinfo {author} {\bibfnamefont {T.~J.~} \bibnamefont {Kippenberg}},\
  }\bibfield  {title} {\enquote {\bibinfo {title} {{Frequency comb assisted
  diode laser spectroscopy for measurement of microcavity dispersion}},}\
  }\href {\doibase 10.1038/nphoton.2009.138} {\bibfield  {journal} {\bibinfo
  {journal} {Nat.~Photonics}\ }\textbf {\bibinfo {volume} {3}},\ \bibinfo
  {pages} {529--533} (\bibinfo {year} {2009})},\   \BibitemShut {NoStop}%
\bibitem [{\citenamefont {Chembo}\ and\ \citenamefont {Yu}(2010)}]{Chembo2010}%
  \BibitemOpen
  \bibfield  {author} {\bibinfo {author} {\bibfnamefont {Y.~K.~}
  \bibnamefont {Chembo}}\ and\ \bibinfo {author} {\bibfnamefont
  {N.~}\bibnamefont {Yu}},\ }\bibfield  {title} {\enquote {\bibinfo {title}
  {{Modal expansion approach to optical-frequency-comb generation with
  monolithic whispering-gallery-mode resonators}},}\ }\href {\doibase
  10.1103/PhysRevA.82.033801} {\bibfield  {journal} {\bibinfo  {journal}
  {Phys.~Rev.~A}\ }\textbf {\bibinfo {volume} {82}},\ \bibinfo {pages}
  {033801} (\bibinfo {year} {2010})}\BibitemShut {NoStop}%
\bibitem [{\citenamefont {Soltani}\ \emph {et~al.}(2012)\citenamefont
  {Soltani}, \citenamefont {Yegnanarayanan}, \citenamefont {Li}, \citenamefont
  {Eftekhar},\ and\ \citenamefont {Adibi}}]{Soltani2012}%
  \BibitemOpen
  \bibfield  {author} {\bibinfo {author} {\bibfnamefont {M.~}
  \bibnamefont {Soltani}}, \bibinfo {author} {\bibfnamefont {S.~}
  \bibnamefont {Yegnanarayanan}}, \bibinfo {author} {\bibfnamefont {Q.~}
  \bibnamefont {Li}}, \bibinfo {author} {\bibfnamefont {A.~A.~} \bibnamefont
  {Eftekhar}}, \ and\ \bibinfo {author} {\bibfnamefont {A.~}\bibnamefont
  {Adibi}},\ }\bibfield  {title} {\enquote {\bibinfo {title} {{Self-sustained
  gigahertz electronic oscillations in ultrahigh-\emph{Q} photonic microresonators}},}\ }\href {\doibase
  10.1103/PhysRevA.85.053819} {\bibfield  {journal} {\bibinfo  {journal}
  {Phys.~Rev.~A}\ }\textbf {\bibinfo {volume} {85}},\ \bibinfo {pages}
  {053819} (\bibinfo {year} {2012})}\BibitemShut {NoStop}%
\bibitem [{\citenamefont {Cazier}\ \emph {et~al.}(2013)\citenamefont {Cazier},
  \citenamefont {Checoury}, \citenamefont {Haret},\ and\ \citenamefont
  {Boucaud}}]{Cazier2013}%
  \BibitemOpen
  \bibfield  {author} {\bibinfo {author} {\bibfnamefont {N.~} \bibnamefont
  {Cazier}}, \bibinfo {author} {\bibfnamefont {X.~} \bibnamefont
  {Checoury}}, \bibinfo {author} {\bibfnamefont {L.-D.~} \bibnamefont
  {Haret}}, \ and\ \bibinfo {author} {\bibfnamefont {P.~} \bibnamefont
  {Boucaud}},\ }\bibfield  {title} {\enquote {\bibinfo {title} {{High-frequency
  self-induced oscillations in a silicon nanocavity}},}\ }\href {\doibase
  10.1364/OE.21.013626} {\bibfield  {journal} {\bibinfo  {journal} {Opt.~
  Express}\ }\textbf {\bibinfo {volume} {21}},\ \bibinfo {pages} {13626}
  (\bibinfo {year} {2013})}\BibitemShut {NoStop}%
\bibitem [{\citenamefont {Pasquazi}\ \emph {et~al.}(2017)\citenamefont
  {Pasquazi}, \citenamefont {Peccianti}, \citenamefont {Razzari}, \citenamefont
  {Moss}, \citenamefont {Coen}, \citenamefont {Erkintalo}, \citenamefont
  {Chembo}, \citenamefont {Hansson}, \citenamefont {Wabnitz}, \citenamefont
  {Del'Haye}, \citenamefont {Xue}, \citenamefont {Weiner},\ and\ \citenamefont
  {Morandotti}}]{Pasquazi2017}%
  \BibitemOpen
  \bibfield  {author} {\bibinfo {author} {\bibfnamefont {A.~} \bibnamefont
  {Pasquazi}}, \bibinfo {author} {\bibfnamefont {M.~} \bibnamefont
  {Peccianti}}, \bibinfo {author} {\bibfnamefont {L.~} \bibnamefont
  {Razzari}}, \bibinfo {author} {\bibfnamefont {D.~J.~} \bibnamefont
  {Moss}}, \bibinfo {author} {\bibfnamefont {S.~} \bibnamefont
  {Coen}}, \bibinfo {author} {\bibfnamefont {M.~} \bibnamefont {Erkintalo}},
  \bibinfo {author} {\bibfnamefont {Y.~K.~} \bibnamefont {Chembo}}, \bibinfo
  {author} {\bibfnamefont {T.~} \bibnamefont {Hansson}}, \bibinfo {author}
  {\bibfnamefont {S.~} \bibnamefont {Wabnitz}}, \bibinfo {author}
  {\bibfnamefont {P.~} \bibnamefont {Del'Haye}}, \bibinfo {author}
  {\bibfnamefont {X.~} \bibnamefont {Xue}}, \bibinfo {author}
  {\bibfnamefont {A.~M.} \bibnamefont {Weiner}}, \ and\ \bibinfo {author}
  {\bibfnamefont {R.~} \bibnamefont {Morandotti}},\ }\bibfield  {title}
  {\enquote {\bibinfo {title} {{Micro-combs: A novel generation of optical
  sources}},}\ }\href {\doibase 10.1016/j.physrep.2017.08.004} {\bibfield
  {journal} {\bibinfo  {journal} {Phys.~Rep.~} } (\bibinfo {year}
  {2017})}\BibitemShut {NoStop}%
\bibitem [{\citenamefont {Herr}\ \emph {et~al.}(2014)\citenamefont {Herr},
  \citenamefont {Brasch}, \citenamefont {Jost}, \citenamefont {Wang},
  \citenamefont {Kondratiev}, \citenamefont {Gorodetsky},\ and\ \citenamefont
  {Kippenberg}}]{Herr2014a}%
  \BibitemOpen
  \bibfield  {author} {\bibinfo {author} {\bibfnamefont {T.~} \bibnamefont
  {Herr}}, \bibinfo {author} {\bibfnamefont {V}~\bibnamefont {Brasch}},
  \bibinfo {author} {\bibfnamefont {J.~D.~} \bibnamefont {Jost}}, \bibinfo
  {author} {\bibfnamefont {C.~Y.~} \bibnamefont {Wang}}, \bibinfo {author}
  {\bibfnamefont {N.~M.~} \bibnamefont {Kondratiev}}, \bibinfo {author}
  {\bibfnamefont {M.~L~.} \bibnamefont {Gorodetsky}}, \ and\ \bibinfo {author}
  {\bibfnamefont {T.~J.~} \bibnamefont {Kippenberg}},\ }\bibfield  {title}
  {\enquote {\bibinfo {title} {{Temporal solitons in optical
  microresonators}},}\ }\href {\doibase 10.1038/nphoton.2013.343} {\bibfield
  {journal} {\bibinfo  {journal} {Nat.~Photonics}\ }\textbf {\bibinfo {volume}
  {8}},\ \bibinfo {pages} {145--152} (\bibinfo {year} {2014})},\  \BibitemShut {NoStop}%
\bibitem [{\citenamefont {Hausmann}\ \emph {et~al.}(2014)\citenamefont
  {Hausmann}, \citenamefont {Bulu}, \citenamefont {Venkataraman}, \citenamefont
  {Deotare}, \citenamefont {Loncar},\ and\ \citenamefont
  {Lon{\v{c}}ar}}]{Hausmann2014}%
  \BibitemOpen
  \bibfield  {author} {\bibinfo {author} {\bibfnamefont {B.~J.~M.}
  \bibnamefont {Hausmann}}, \bibinfo {author} {\bibfnamefont {I.~}\bibnamefont
  {Bulu}}, \bibinfo {author} {\bibfnamefont {V.~}\bibnamefont {Venkataraman}},
  \bibinfo {author} {\bibfnamefont {P.~}\bibnamefont {Deotare}}, \bibinfo
  {author} {\bibfnamefont {M.}~\bibnamefont {Loncar}}, \ and\ \bibinfo {author}
  {\bibfnamefont {M.}~\bibnamefont {Lon{\v{c}}ar}},\ }\bibfield  {title}
  {\enquote {\bibinfo {title} {{Diamond nonlinear photonics}},}\ }\href
  {\doibase 10.1038/nphoton.2014.72} {\bibfield  {journal} {\bibinfo  {journal}
  {Nat.~Photonics}\ }\textbf {\bibinfo {volume} {8}},\ \bibinfo {pages}
  {369--374} (\bibinfo {year} {2014})}\BibitemShut {NoStop}%
\bibitem [{\citenamefont {Grinberg}\ \emph {et~al.}(2012)\citenamefont
  {Grinberg}, \citenamefont {Bencheikh}, \citenamefont {Brunstein},
  \citenamefont {Yacomotti}, \citenamefont {Dumeige}, \citenamefont {Sagnes},
  \citenamefont {Raineri}, \citenamefont {Bigot},\ and\ \citenamefont
  {Levenson}}]{Grinberg2012}%
  \BibitemOpen
  \bibfield  {author} {\bibinfo {author} {\bibfnamefont {P.}~\bibnamefont
  {Grinberg}}, \bibinfo {author} {\bibfnamefont {K.}~\bibnamefont {Bencheikh}},
  \bibinfo {author} {\bibfnamefont {M.}~\bibnamefont {Brunstein}}, \bibinfo
  {author} {\bibfnamefont {A.~M.}\ \bibnamefont {Yacomotti}}, \bibinfo {author}
  {\bibfnamefont {Y.~} \bibnamefont {Dumeige}}, \bibinfo {author}
  {\bibfnamefont {I.~} \bibnamefont {Sagnes}}, \bibinfo {author}
  {\bibfnamefont {F.}~\bibnamefont {Raineri}}, \bibinfo {author} {\bibfnamefont
  {L.}~\bibnamefont {Bigot}}, \ and\ \bibinfo {author} {\bibfnamefont {J.~A.~}
  \bibnamefont {Levenson}},\ }\bibfield  {title} {\enquote {\bibinfo {title}
  {{Nanocavity Linewidth Narrowing and Group Delay Enhancement by Slow Light
  Propagation and Nonlinear Effects}},}\ }\href {\doibase
  10.1103/PhysRevLett.109.113903} {\bibfield  {journal} {\bibinfo  {journal}
  {Phys.~Rev.~Lett.~} }\textbf {\bibinfo {volume} {109}},\ \bibinfo
  {pages} {113903} (\bibinfo {year} {2012})}\BibitemShut {NoStop}%
\bibitem [{\citenamefont {Faraon}\ \emph {et~al.}(2010)\citenamefont {Faraon},
  \citenamefont {Barclay}, \citenamefont {Santori}, \citenamefont {Fu},\ and\
  \citenamefont {Beausoleil}}]{Faraon2010a}%
  \BibitemOpen
  \bibfield  {author} {\bibinfo {author} {\bibfnamefont {A.~} \bibnamefont
  {Faraon}}, \bibinfo {author} {\bibfnamefont {P.~E.~} \bibnamefont
  {Barclay}}, \bibinfo {author} {\bibfnamefont {C. ~} \bibnamefont
  {Santori}}, \bibinfo {author} {\bibfnamefont {K.-M.~C.}\ \bibnamefont
  {Fu}}, \ and\ \bibinfo {author} {\bibfnamefont {Raymond~G.}\ \bibnamefont
  {Beausoleil}},\ }\bibfield  {title} {\enquote {\bibinfo {title} {{Resonant
  enhancement of the zero-phonon emission from a color center in a diamond
  cavity}},}\ }\href {\doibase 10.1038/nphoton.2011.52} {\bibfield  {journal}
  {\bibinfo  {journal} {Nat.~Photonics}\ }\textbf {\bibinfo {volume} {5}},\
  \bibinfo {pages} {301--305} (\bibinfo {year} {2010})},\  \BibitemShut {NoStop}%
\bibitem [{\citenamefont {Malaguti}\ \emph {et~al.}(2011)\citenamefont
  {Malaguti}, \citenamefont {Bellanca}, \citenamefont {{De Rossi}},
  \citenamefont {Combri{\'{e}}},\ and\ \citenamefont {Trillo}}]{Malaguti2011c}%
  \BibitemOpen
  \bibfield  {author} {\bibinfo {author} {\bibfnamefont {S.~}
  \bibnamefont {Malaguti}}, \bibinfo {author} {\bibfnamefont {G.~}
  \bibnamefont {Bellanca}}, \bibinfo {author} {\bibfnamefont {A.~}
  \bibnamefont {{De Rossi}}}, \bibinfo {author} {\bibfnamefont {S.~}
  \bibnamefont {Combri{\'e}}}, \ and\ \bibinfo {author} {\bibfnamefont
  {S.~} \bibnamefont {Trillo}},\ }\bibfield  {title} {\enquote {\bibinfo
  {title} {{Self-pulsing driven by two-photon absorption in semiconductor
  nanocavities}},}\ }\href {\doibase 10.1103/PhysRevA.83.051802} {\bibfield
  {journal} {\bibinfo  {journal} {Phys.~Rev.~A}\ }\textbf {\bibinfo
  {volume} {83}},\ \bibinfo {pages} {051802} (\bibinfo {year}
  {2011})}\BibitemShut {NoStop}%
\bibitem [{\citenamefont {Armaroli}\ \emph {et~al.}(2011)\citenamefont
  {Armaroli}, \citenamefont {Malaguti}, \citenamefont {Bellanca}, \citenamefont
  {Trillo}, \citenamefont {de~Rossi},\ and\ \citenamefont
  {Combri{\'{e}}}}]{Armaroli2011a}%
  \BibitemOpen
  \bibfield  {author} {\bibinfo {author} {\bibfnamefont {A.~} \bibnamefont
  {Armaroli}}, \bibinfo {author} {\bibfnamefont {S. ~} \bibnamefont
  {Malaguti}}, \bibinfo {author} {\bibfnamefont {G. ~} \bibnamefont
  {Bellanca}}, \bibinfo {author} {\bibfnamefont {Stefano}\ \bibnamefont
  {Trillo}}, \bibinfo {author} {\bibfnamefont {A.~} \bibnamefont
  {de~Rossi}}, \ and\ \bibinfo {author} {\bibfnamefont {S.~} \bibnamefont
  {Combri{\'{e}}}},\ }\bibfield  {title} {\enquote {\bibinfo {title}
  {{Oscillatory dynamics in nanocavities with noninstantaneous Kerr
  response}},}\ }\href {\doibase 10.1103/PhysRevA.84.053816} {\bibfield
  {journal} {\bibinfo  {journal} {Phys.~Rev.~A}\ }\textbf {\bibinfo
  {volume} {84}},\ \bibinfo {pages} {053816} (\bibinfo {year}
  {2011})}\BibitemShut {NoStop}%
\bibitem [{\citenamefont {Vaerenbergh}\ \emph {et~al.}(2012)\citenamefont
  {Vaerenbergh}, \citenamefont {Fiers}, \citenamefont {Dambre},\ and\
  \citenamefont {Bienstman}}]{Vaerenbergh2012}%
  \BibitemOpen
  \bibfield  {author} {\bibinfo {author} {\bibfnamefont {T.~}
  \bibnamefont {Van Vaerenbergh}}, \bibinfo {author} {\bibfnamefont {M.~}
  \bibnamefont {Fiers}}, \bibinfo {author} {\bibfnamefont {J.~} \bibnamefont
  {Dambre}}, \ and\ \bibinfo {author} {\bibfnamefont {P.~} \bibnamefont
  {Bienstman}},\ }\bibfield  {title} {\enquote {\bibinfo {title} {{Simplified
  description of self-pulsation and excitability by thermal and free-carrier
  effects in semiconductor microcavities}},}\ }\href {\doibase
  10.1103/PhysRevA.86.063808} {\bibfield  {journal} {\bibinfo  {journal}
  {Phys.~Rev.~A}\ }\textbf {\bibinfo {volume} {86}},\ \bibinfo {pages}
  {063808} (\bibinfo {year} {2012})}\BibitemShut {NoStop}%
\bibitem [{\citenamefont {Huet}\ \emph {et~al.}(2016)\citenamefont {Huet},
  \citenamefont {Rasoloniaina}, \citenamefont {Guillem{\'{e}}}, \citenamefont
  {Rochard}, \citenamefont {F{\'{e}}ron}, \citenamefont {Mortier},
  \citenamefont {Levenson}, \citenamefont {Bencheikh}, \citenamefont
  {Yacomotti},\ and\ \citenamefont {Dumeige}}]{Huet2016}%
  \BibitemOpen
  \bibfield  {author} {\bibinfo {author} {\bibfnamefont {V.~} \bibnamefont
  {Huet}}, \bibinfo {author} {\bibfnamefont {A.}~\bibnamefont {Rasoloniaina}},
  \bibinfo {author} {\bibfnamefont {P.}~\bibnamefont {Guillem{\'{e}}}},
  \bibinfo {author} {\bibfnamefont {P.}~\bibnamefont {Rochard}}, \bibinfo
  {author} {\bibfnamefont {P.}~\bibnamefont {F{\'{e}}ron}}, \bibinfo {author}
  {\bibfnamefont {M.}~\bibnamefont {Mortier}}, \bibinfo {author} {\bibfnamefont
  {A.}~\bibnamefont {Levenson}}, \bibinfo {author} {\bibfnamefont
  {K.}~\bibnamefont {Bencheikh}}, \bibinfo {author} {\bibfnamefont
  {A.}~\bibnamefont {Yacomotti}}, \ and\ \bibinfo {author} {\bibfnamefont
  {Y.~} \bibnamefont {Dumeige}},\ }\bibfield  {title} {\enquote {\bibinfo
  {title} {{Millisecond photon lifetime in a slow-light microcavity}},}\ }\href
  {\doibase 10.1103/PhysRevLett.116.133902} {\bibfield  {journal} {\bibinfo
  {journal} {Phys.~Rev.~Lett.~} }\textbf {\bibinfo {volume} {116}},\
  \bibinfo {pages} {133902} (\bibinfo {year} {2016})}\BibitemShut {NoStop}%
\bibitem [{\citenamefont {Maes}\ \emph {et~al.}(2009)\citenamefont {Maes},
  \citenamefont {Fiers},\ and\ \citenamefont {Bienstman}}]{Maes2009}%
  \BibitemOpen
  \bibfield  {author} {\bibinfo {author} {\bibfnamefont {B.~}
  \bibnamefont {Maes}}, \bibinfo {author} {\bibfnamefont {M.~} \bibnamefont
  {Fiers}}, \ and\ \bibinfo {author} {\bibfnamefont {P.~} \bibnamefont
  {Bienstman}},\ }\bibfield  {title} {\enquote {\bibinfo {title} {{Self-pulsing
  and chaos in short chains of coupled nonlinear microcavities}},}\ }\href
  {\doibase 10.1103/PhysRevA.80.033805} {\bibfield  {journal} {\bibinfo
  {journal} {Phys.~Rev.~A}\ }\textbf {\bibinfo {volume} {80}},\ \bibinfo
  {pages} {033805} (\bibinfo {year} {2009})}\BibitemShut {NoStop}%
\bibitem [{\citenamefont {Grigoriev}\ and\ \citenamefont
  {Biancalana}(2011)}]{Grigoriev2011a}%
  \BibitemOpen
  \bibfield  {author} {\bibinfo {author} {\bibfnamefont {V.~}\bibnamefont
  {Grigoriev}}\ and\ \bibinfo {author} {\bibfnamefont {F.~}\bibnamefont
  {Biancalana}},\ }\bibfield  {title} {\enquote {\bibinfo {title} {{Resonant
  self-pulsations in coupled nonlinear microcavities}},}\ }\href {\doibase
  10.1103/PhysRevA.83.043816} {\bibfield  {journal} {\bibinfo  {journal}
  {Phys.~Rev.~A}\ }\textbf {\bibinfo {volume} {83}},\ \bibinfo {pages}
  {043816} (\bibinfo {year} {2011})},\  \BibitemShut {NoStop}%
\bibitem [{\citenamefont {Dumeige}\ and\ \citenamefont
  {F{\'{e}}ron}(2015)}]{Dumeige2015}%
  \BibitemOpen
  \bibfield  {author} {\bibinfo {author} {\bibfnamefont {Y.~}\bibnamefont
  {Dumeige}}\ and\ \bibinfo {author} {\bibfnamefont {P.~} \bibnamefont
  {F{\'{e}}ron}},\ }\bibfield  {title} {\enquote {\bibinfo {title} {{Coupled
  optical microresonators for microwave all-optical generation and
  processing}},}\ }\href {\doibase 10.1364/OL.40.003237} {\bibfield  {journal}
  {\bibinfo  {journal} {Opt.~Lett.~} }\textbf {\bibinfo {volume} {40}},\
  \bibinfo {pages} {3237--3240} (\bibinfo {year} {2015})}\BibitemShut {NoStop}%
\bibitem{Milian2018} C. Mili{\'a}n, Y. V. Kartashov, D. V. Skryabin, and L. Torner, ``Cavity solitons in a microring dimer with gain and loss''  \href {\doibase 10.1364/OL.43.000979}{Opt. Lett., \textbf{43}, 979, (2018)}
\bibitem [{\citenamefont {Armaroli}\ \emph {et~al.}(2015)\citenamefont
  {Armaroli}, \citenamefont {Feron},\ and\ \citenamefont
  {Dumeige}}]{Armaroli2015cav}%
  \BibitemOpen
  \bibfield  {author} {\bibinfo {author} {\bibfnamefont {A.~} \bibnamefont
  {Armaroli}}, \bibinfo {author} {\bibfnamefont {P.}\ \bibnamefont
  {F{\'e}ron}}, \ and\ \bibinfo {author} {\bibfnamefont {Y.~} \bibnamefont
  {Dumeige}},\ }\bibfield  {title} {\enquote {\bibinfo {title} {{Stable
  integrated hyper-parametric oscillator based on coupled optical
  microcavities}},}\ }\href {\doibase 10.1364/OL.40.005622} {\bibfield
  {journal} {\bibinfo  {journal} {Opt.~Lett.~} }\textbf {\bibinfo {volume}
  {40}},\ \bibinfo {pages} {5622--5625} (\bibinfo {year} {2015})}\BibitemShut
  {NoStop}%
\bibitem [{\citenamefont {Belotti}\ \emph {et~al.}(2010)\citenamefont
  {Belotti}, \citenamefont {Galli}, \citenamefont {Gerace}, \citenamefont
  {Andreani}, \citenamefont {Guizzetti}, \citenamefont {{Md Zain}},
  \citenamefont {Johnson}, \citenamefont {Sorel},\ and\ \citenamefont {{De La
  Rue}}}]{Belotti2010a}%
  \BibitemOpen
  \bibfield  {author} {\bibinfo {author} {\bibfnamefont {M.~}
  {Belotti}}, \bibinfo {author} {\bibfnamefont {M.~} \bibnamefont {Galli}},
  \bibinfo {author} {\bibfnamefont {D.~} \bibnamefont {Gerace}}, \bibinfo
  {author} {\bibfnamefont {L.~C.~} \bibnamefont {Andreani}}, \bibinfo
  {author} {\bibfnamefont {G.~} \bibnamefont {Guizzetti}}, \bibinfo
  {author} {\bibfnamefont {A.~R.~} \bibnamefont {{Md Zain}}}, \bibinfo
  {author} {\bibfnamefont {N.~P.~} \bibnamefont {Johnson}}, \bibinfo
  {author} {\bibfnamefont {M.~} \bibnamefont {Sorel}}, \ and\ \bibinfo
  {author} {\bibfnamefont {R.~M.}\ \bibnamefont {{De La Rue}}},\
  }\bibfield  {title} {\enquote {\bibinfo {title} {{All-optical switching in
  silicon-on-insulator photonic wire nano-cavities}},}\ }\href {\doibase
  10.1364/OE.18.001450} {\bibfield  {journal} {\bibinfo  {journal} {Opt.~Express}\ }\textbf {\bibinfo {volume} {18}},\ \bibinfo {pages} {1450--1461}
  (\bibinfo {year} {2010})}\BibitemShut {NoStop}%
\bibitem [{\citenamefont {Hansson}\ \emph {et~al.}(2013)\citenamefont
  {Hansson}, \citenamefont {Modotto},\ and\ \citenamefont
  {Wabnitz}}]{Hansson2013}%
  \BibitemOpen
  \bibfield  {author} {\bibinfo {author} {\bibfnamefont {T.~} \bibnamefont
  {Hansson}}, \bibinfo {author} {\bibfnamefont {D.~}\bibnamefont {Modotto}}, \
  and\ \bibinfo {author} {\bibfnamefont {S.~} \bibnamefont {Wabnitz}},\
  }\bibfield  {title} {\enquote {\bibinfo {title} {{Dynamics of the
  modulational instability in microresonator frequency combs}},}\ }\href
  {\doibase 10.1103/PhysRevA.88.023819} {\bibfield  {journal} {\bibinfo
  {journal} {Phys.~Rev.~A}\ }\textbf {\bibinfo {volume} {88}},\ \bibinfo
  {pages} {023819} (\bibinfo {year} {2013})}.\  \BibitemShut {NoStop}%
\bibitem [{\citenamefont {Strekalov}\ and\ \citenamefont
  {Yu}(2009)}]{Strekalov2009}%
  \BibitemOpen
  \bibfield  {author} {\bibinfo {author} {\bibfnamefont {D.~V.~}\
  \bibnamefont {Strekalov}}\ and\ \bibinfo {author} {\bibfnamefont {N.~}\
  \bibnamefont {Yu}},\ }\bibfield  {title} {\enquote {\bibinfo {title}
  {{Generation of optical combs in a whispering gallery mode resonator from a
  bichromatic pump}},}\ }\href {\doibase 10.1103/PhysRevA.79.041805} {\bibfield
   {journal} {\bibinfo  {journal} {Phys.~Rev.~A}\ }\textbf {\bibinfo {volume} {79}},\ \bibinfo {pages} {041805(R)}
  (\bibinfo {year} {2009})},\  \BibitemShut {NoStop}%
\bibitem [{\citenamefont {Hansson}\ and\ \citenamefont
  {Wabnitz}(2014)}]{Hansson2014d}%
  \BibitemOpen
  \bibfield  {author} {\bibinfo {author} {\bibfnamefont {T.~}\bibnamefont
  {Hansson}}\ and\ \bibinfo {author} {\bibfnamefont {S.~}\bibnamefont
  {Wabnitz}},\ }\bibfield  {title} {\enquote {\bibinfo {title}
  {{Bichromatically pumped microresonator frequency combs}},}\ }\href {\doibase
  10.1103/PhysRevA.90.013811} {\bibfield  {journal} {\bibinfo  {journal}
  {Phys.~Rev.~A}\ }\textbf
  {\bibinfo {volume} {90}},\ \bibinfo {pages} {013811} (\bibinfo {year}
  {2014})},\ 
  \BibitemShut {NoStop}%
\bibitem [{\citenamefont {Haus}(1983)}]{HausBook}%
  \BibitemOpen
  \bibfield  {author} {\bibinfo {author} {\bibfnamefont {H.~A.~}
  \bibnamefont {Haus}},\ }\href@noop {} {\emph {\bibinfo {title} {{Waves and
  Fields in Optoelectronics}}}}\ (\bibinfo  {publisher} {Prentice-Hall series
  in solid state physical electronics},\ \bibinfo {year} {1983})\ p.\ \bibinfo
  {pages} {402}\BibitemShut {NoStop}%
\bibitem [{\citenamefont {Fan}\ \emph {et~al.}(2003)\citenamefont {Fan},
  \citenamefont {Suh},\ and\ \citenamefont {Joannopoulos}}]{Fan2003}%
  \BibitemOpen
  \bibfield  {author} {\bibinfo {author} {\bibfnamefont {S.~}\bibnamefont
  {Fan}}, \bibinfo {author} {\bibfnamefont {W.~}\bibnamefont {Suh}}, \
  and\ \bibinfo {author} {\bibfnamefont {J.~D.~}\bibnamefont
  {Joannopoulos}},\ }\bibfield  {title} {\enquote {\bibinfo {title} {{Temporal
  coupled-mode theory for the Fano resonance in optical resonators}},}\ }\href
  {\doibase 10.1364/JOSAA.20.000569} {\bibfield  {journal} {\bibinfo  {journal}
  {J.~Opt.~Soc.~Am.~A}\ }\textbf {\bibinfo {volume}
  {20}},\ \bibinfo {pages} {569--572} (\bibinfo {year} {2003})}\BibitemShut
  {NoStop}%
\bibitem [{\citenamefont {Abdollahi}\ and\ \citenamefont
  {Van}(2014)}]{Abdollahi2014}%
  \BibitemOpen
  \bibfield  {author} {\bibinfo {author} {\bibfnamefont {S.~}\bibnamefont
  {Abdollahi}}\ and\ \bibinfo {author} {\bibfnamefont {V.~}\bibnamefont
  {Van}},\ }\bibfield  {title} {\enquote {\bibinfo {title} {{Analysis of
  optical instability in coupled microring resonators}},}\ }\href {\doibase
  10.1364/JOSAB.31.003081} {\bibfield  {journal} {\bibinfo  {journal} {J.Opt.~Soc.~Am.~B }}\textbf {\bibinfo {volume} {31}},\
  \bibinfo {pages} {3081--3087} (\bibinfo {year} {2014})}\BibitemShut {NoStop}%
  \bibitem{Alphonse2014} A. Rasoloniaina, V. Huet, T. K. N. Nguyên, E. Le Cren, M. Mortier, L. Michely, Y. Dumeige, and P. Féron, ``Controling the coupling properties of active ultrahigh-Q WGM microcavities from undercoupling to selective amplification,'' \href{{\doibase 10.1038/srep04023}}{Sci. Rep. \textbf{4},  4023, 2014.}
\bibitem [{\citenamefont {Sracic}\ and\ \citenamefont
  {Allen}(2011)}]{Sracic2011}%
  \BibitemOpen
  \bibfield  {author} {\bibinfo {author} {\bibfnamefont {M.~W.~}
  \bibnamefont {Sracic}}\ and\ \bibinfo {author} {\bibfnamefont {M.~S.~}
  \bibnamefont {Allen}},\ }\bibfield  {title} {\enquote {\bibinfo {title}
  {{Numerical Continuation of Periodic Orbits for Harmonically Forced Nonlinear
  Systems}},}\ }in\ \href {\doibase 10.1007/978-1-4419-9316-8_5} {\emph
  {\bibinfo {booktitle} {Civil Engineering Topics, Volume 4 Conference
  Proceedings of the Society for Experimental Mechanics Series. Springer, New
  York, NY}}},\ \bibinfo {editor} {edited by\ \bibinfo {editor} {\bibfnamefont
  {Proulx}\ \bibnamefont {T.}}}\ (\bibinfo  {publisher} {Spinger},\ \bibinfo
  {address} {New York},\ \bibinfo {year} {2011})\  \bibinfo {pages}
  {51--69}\BibitemShut {NoStop}%
\bibitem [{\citenamefont {Dhooge}\ \emph {et~al.}(2003)\citenamefont {Dhooge},
  \citenamefont {Govaerts},\ and\ \citenamefont {Kuznetsov}}]{Dhooge2003}%
  \BibitemOpen
  \bibfield  {author} {\bibinfo {author} {\bibfnamefont {A.}~\bibnamefont
  {Dhooge}}, \bibinfo {author} {\bibfnamefont {W.}~\bibnamefont {Govaerts}}, \
  and\ \bibinfo {author} {\bibfnamefont {Yu.~A.}\ \bibnamefont {Kuznetsov}},\
  }\bibfield  {title} {\enquote {\bibinfo {title} {{MATCONT: A MATLAB package
  for numerical bifurcation analysis of ODEs}},}\ }\href {\doibase
  10.1145/779359.779362} {\bibfield  {journal} {\bibinfo  {journal} {ACM
  Transactions on Mathematical Software}\ }\textbf {\bibinfo {volume} {29}},\
  \bibinfo {pages} {141--164} (\bibinfo {year} {2003})}\BibitemShut {NoStop}%
\bibitem [{\citenamefont {Trillo}\ \emph {et~al.}(1994)\citenamefont {Trillo},
  \citenamefont {Wabnitz},\ and\ \citenamefont {Kennedy}}]{Trillo1994}%
  \BibitemOpen
  \bibfield  {author} {\bibinfo {author} {\bibfnamefont {S.~}\bibnamefont
  {Trillo}}, \bibinfo {author} {\bibfnamefont {S.~}\bibnamefont
  {Wabnitz}}, \ and\ \bibinfo {author} {\bibfnamefont {T.~A.~B.~}
  \bibnamefont {Kennedy}},\ }\bibfield  {title} {\enquote {\bibinfo {title}
  {{Nonlinear dynamics of dual-frequency-pumped multiwave mixing in optical
  fibers}},}\ }\href {\doibase 10.1103/PhysRevA.50.1732} {\bibfield  {journal}
  {\bibinfo  {journal} {Phys.~Rev.~A}\ }\textbf {\bibinfo {volume} {50}},\
  \bibinfo {pages} {1732--1747} (\bibinfo {year} {1994})}\BibitemShut {NoStop}%
\bibitem [{\citenamefont {Marhic}(2007)}]{MarhicBook2007}%
  \BibitemOpen
  \bibfield  {author} {\bibinfo {author} {\bibfnamefont {M.~E.~}\
  \bibnamefont {Marhic}},\ }\href
  {https://www.amazon.com/Optical-Parametric-Amplifiers-Oscillators-Related/dp/0521861020?SubscriptionId=0JYN1NVW651KCA56C102{\&}tag=techkie-20{\&}linkCode=xm2{\&}camp=2025{\&}creative=165953{\&}creativeASIN=0521861020}
  {\emph {\bibinfo {title} {{Fiber Optical Parametric Amplifiers, Oscillators
  and Related Devices}}}}\ (\bibinfo  {publisher} {Cambridge University
  Press},\ \bibinfo {year} {2007})\BibitemShut {NoStop}%
\bibitem [{\citenamefont {Ott}\ \emph {et~al.}(2013)\citenamefont {O.~},
  \citenamefont {Steffensen}, \citenamefont {K.~Rottwitt},\ and\ \citenamefont
  {McKinstrie}}]{Ott2013c}%
  \BibitemOpen
  \bibfield  {author} {\bibinfo {author} {\bibfnamefont {J.~R.}\ \bibnamefont
  {Ott}}, \bibinfo {author} {\bibfnamefont {H.}~\bibnamefont {Steffensen}},
  \bibinfo {author} {\bibfnamefont {K.}~\bibnamefont {Rottwitt}}, \ and\
  \bibinfo {author} {\bibfnamefont {C.~J.}\ \bibnamefont {McKinstrie}},\
  }\bibfield  {title} {\enquote {\bibinfo {title} {{Geometric interpretation of
  four-wave mixing}},}\ }\href {\doibase 10.1103/PhysRevA.88.043805} {\bibfield
   {journal} {\bibinfo  {journal} {Phys.~Rev.~A}\ }\textbf {\bibinfo
  {volume} {88}},\ \bibinfo {pages} {043805} (\bibinfo {year}
  {2013})}\BibitemShut {NoStop}%
\bibitem [{\citenamefont {Wagner}\ \emph {et~al.}(2009)\citenamefont {Wagner},
  \citenamefont {Holmes}, \citenamefont {Younis}, \citenamefont {Helmy},
  \citenamefont {Hutchings},\ and\ \citenamefont {Aitchison}}]{Wagner2009}%
  \BibitemOpen
  \bibfield  {author} {\bibinfo {author} {\bibfnamefont {S.~J.~}\bibnamefont
  {Wagner}}, \bibinfo {author} {\bibfnamefont {B.~M.~}\bibnamefont
  {Holmes}}, \bibinfo {author} {\bibfnamefont {U.~}\bibnamefont {Younis}},
  \bibinfo {author} {\bibfnamefont {A.~S.}\ \bibnamefont {Helmy}}, \bibinfo
  {author} {\bibfnamefont {David~C.}\ \bibnamefont {Hutchings}}, \ and\
  \bibinfo {author} {\bibfnamefont {J.~S.~}\bibnamefont {Aitchison}},\
  }\bibfield  {title} {\enquote {\bibinfo {title} {{Controlling third-order
  nonlinearities by ion-implantation quantum-well intermixing}},}\ }\href
  {\doibase 10.1109/LPT.2008.2008629} {\bibfield  {journal} {\bibinfo
  {journal} {IEEE Photonics Tech.~Lett.}\ }\textbf {\bibinfo {volume}
  {21}},\ \bibinfo {pages} {85--87} (\bibinfo {year} {2009})}\BibitemShut
  {NoStop}%
\bibitem [{\citenamefont {Little}\ \emph {et~al.}(1997)\citenamefont {Little},
  \citenamefont {Chu}, \citenamefont {Haus}, \citenamefont {Foresi},\ and\
  \citenamefont {Laine}}]{Little1997}%
  \BibitemOpen
  \bibfield  {author} {\bibinfo {author} {\bibfnamefont {B.~E.~}\bibnamefont
  {Little}}, \bibinfo {author} {\bibfnamefont {S~T}\ \bibnamefont {Chu}},
  \bibinfo {author} {\bibfnamefont {H.~A.~}\bibnamefont {Haus}}, \bibinfo
  {author} {\bibfnamefont {J.~}~\bibnamefont {Foresi}}, \ and\ \bibinfo {author}
  {\bibfnamefont {J.-P.~}\bibnamefont {Laine}},\ }\bibfield  {title} {\enquote
  {\bibinfo {title} {{Microring resonator channel dropping filters}},}\ }  
  {\bibfield  {journal} {\bibinfo  {journal} {J.~Lightwave Tech.~}}\textbf {\bibinfo {volume} {15}},\ \bibinfo {pages} {998--1005}
  (\bibinfo {year} {1997})}\BibitemShut {NoStop}%
\bibitem [{\citenamefont {Okawachi}\ \emph {et~al.}(2015)\citenamefont
  {Okawachi}, \citenamefont {Yu}, \citenamefont {Luke}, \citenamefont
  {Carvalho}, \citenamefont {Ramelow}, \citenamefont {Farsi}, \citenamefont
  {Lipson},\ and\ \citenamefont {Gaeta}}]{Okawachi2015}%
  \BibitemOpen
  \bibfield  {author} {\bibinfo {author} {\bibfnamefont {Y.}
  \bibnamefont {Okawachi}}, \bibinfo {author} {\bibfnamefont {M.}
  \bibnamefont {Yu}}, \bibinfo {author} {\bibfnamefont {K.}\bibnamefont
  {Luke}}, \bibinfo {author} {\bibfnamefont {D.~O.} \bibnamefont
  {Carvalho}}, \bibinfo {author} {\bibfnamefont {S.} \bibnamefont
  {Ramelow}}, \bibinfo {author} {\bibfnamefont {A.} \bibnamefont
  {Farsi}}, \bibinfo {author} {\bibfnamefont {M.} \bibnamefont {Lipson}},
  \ and\ \bibinfo {author} {\bibfnamefont {A.~L.} \bibnamefont
  {Gaeta}},\ }\bibfield  {title} {\enquote {\bibinfo {title} {{Dual-pumped
  degenerate Kerr oscillator in a silicon nitride microresonator}},}\ }\href
  {\doibase 10.1364/OL.40.005267} {\bibfield  {journal} {\bibinfo  {journal}
  {Opt.~Lett.}\ }\textbf {\bibinfo {volume} {40}},\ \bibinfo {pages} {5267--5270}
  (\bibinfo {year} {2015})}.\

\bibitem{Zheng2014} J. Zheng, H. Wang, W. Li, L. Wang, T. Su, J. Liu, and N. Zhu, ``Photonic-assisted microwave frequency multiplier based on nonlinear polarization rotation,'' \href{\doibase 10.1364/OL.39.001366}{Opt.~Lett.~\textbf{39}, 1366 (2014)}.

\bibitem{Long2016} Y. Long, L. Zhou, and J. Wang, ``Photonic-assisted microwave signal multiplication and modulation using a silicon Mach–Zehnder modulator,'' \href{\doibase 10.1038/srep20215}{ Sci. Rep. \textbf{6}, 20215 (2016)}.
\end{thebibliography}
%

\end{document}